\documentclass[aps,prb,twocolumn,superscriptaddress,floatfix]{revtex4-2}
\usepackage[utf8]{inputenc}
\usepackage{natbib}
\usepackage[colorlinks=true,linkcolor=blue,citecolor=blue,urlcolor=blue]{hyperref}
\usepackage{graphicx}
\usepackage{latexsym}
\usepackage{amssymb}
\usepackage{amsmath}
\usepackage{amsfonts}
\usepackage{amsthm}
\usepackage{bm}
\usepackage{floatrow}
\usepackage[caption=false]{subfig}
\usepackage{bbm}
\usepackage{enumitem}
\usepackage{hyperref}
\usepackage{lipsum}
\usepackage{braket}
\usepackage{tikz}
\usepackage[margin=1in]{geometry}
\usetikzlibrary{shapes.geometric, arrows.meta}
\usepackage{tikzit}

\tikzstyle{blkdot}=[fill=black, draw=black, shape=circle]

\tikzstyle{green}=[draw=red, fill={rgb,255: red,255; green,24; blue,55}, ->]
\tikzstyle{gree}=[-, draw={rgb,255: red,12; green,191; blue,84}]

\usepackage{pgfplots}
\usepackage{comment}
\usepackage{diagbox}
\usepackage{ifthen}
\usepackage{ragged2e}
\usepackage{multirow}
\usepackage[export]{adjustbox}
\usetikzlibrary{arrows,decorations.pathreplacing,decorations.markings,arrows.meta,patterns,3d}
\usepackage[normalem]{ulem} 
\usepackage{cancel}
\usepackage{microtype}
\usepackage{booktabs}

\theoremstyle{definition}

\newcommand{\plaq}{\begin{tikzpicture}[baseline]
  \def\R{1}
  \foreach \i in {1,...,6} {
    \coordinate (V\i) at ({60+60*(\i-1)}:\R);
  }
  \foreach \i [evaluate={\j=int(mod(\i,6)+1);}] in {1,...,6} {
    \draw[thick,red] (V\i) -- (V\j);
  }
\foreach \i in {1,...,6} {
    \draw[dotted] (0,0) -- (V\i);
  }
  \foreach \i in {1,...,6} {
    \fill (V\i) circle (2pt);
  }
\fill (0,0) circle (2pt);
\node[below right] at (0.15,0.15) {$X_v$};
\end{tikzpicture}
}

\newcommand{\DS}{
\begin{tikzpicture}[baseline][scale=1]
    \foreach \i in {0,...,5} {
        \coordinate (v\i) at ({cos(60*\i)},{sin(60*\i)});
    }
    \draw[thick] (v0) -- (v1) -- (v2) -- (v3) -- (v4) -- (v5) -- cycle;
    \foreach \i in {0,...,5} {
        \draw[thick] (v\i) -- ({1.5*cos(60*\i)},{1.5*sin(60*\i)});
    }
    \foreach \i in {0,...,5} {
        \node[font=\small] at ({1.05*cos(60*\i+30)},{1.05*sin(60*\i+30)}) {$X$};
    }
    \node at (0,0) {$p$};
    \foreach \i in {0,...,5} {
        \node[font=\small] at ({1.7*cos(60*\i)},{1.7*sin(60*\i)}) {$S$};
    }
\end{tikzpicture}}

\newcommand{\vertex}{
\begin{tikzpicture}[baseline][scale=1, every node/.style={font=\small}]
    \foreach \angle/\label in {90/1, 210/2, 330/3} {
        \draw[thick]  -- ++(\angle:1) coordinate (e\label);
    }
    \draw[thick] (0,0) -- ++ (0,0.9);
    \draw[thick] (0,0) -- ++ (0.779,-0.45);
    \draw[thick] (0,0) -- ++ (-0.779,-0.45);
    \node at (0.2,0.5) {$Z$};
    \node at (0.5,0) {$Z$};
    \node at (-0.45,0) {$Z$};
\end{tikzpicture}
}
\pgfplotsset{compat=1.8}

\begin{document}


\makeatletter
  \@namedef{figure}{\killfloatstyle\def\@captype{figure}\FR@redefs
    \flrow@setlist{{figure}}%
    \columnwidth\columnwidth\edef\FBB@wd{\the\columnwidth}%
    \FRifFBOX\@@setframe\relax\@@FStrue\@float{figure}}%
\makeatother


\title{Transition between 2D Symmetry Protected Topological Phases on a Klein Bottle}
\date{\today}

\author{Vibhu Ravindran}
\affiliation{Department of Physics and Institute for Quantum Information and Matter, \mbox{California Institute of Technology, Pasadena, CA, 91125, USA}}
\author{Bowen Yang}
\affiliation{Center of Mathematical Sciences and Applications, Harvard University, Cambridge, Massachusetts 02138, USA}
\author{Xie Chen}
\affiliation{Department of Physics and Institute for Quantum Information and Matter, \mbox{California Institute of Technology, Pasadena, CA, 91125, USA}}

\begin{abstract}
Manifolds with nontrivial topology play an essential role in the study of topological phases of matter. In this paper, we study the nontrivial symmetry response of the $2+1$D $Z_2$ symmetry-protected topological (SPT) phase when the system is put on a non-orientable manifold -- the Klein bottle. In particular, we find that when a symmetry defect is inserted along the orientation-reversing cycle of the Klein bottle, the ground state of the system gets an extra charge. This response remains well defined at transition points into the trivial SPT phase, resulting in an exact two-fold degeneracy in the ground state independent of the system size. We demonstrate the symmetry response using exactly solvable lattice models of the SPT phase, as well as numerical work across the transition. We explore the connection of this result to the modular transformation of the $3+1$D $Z_2$ gauge theory and the emergent nature of the parity symmetry in the $Z_2$ SPT phase.
\end{abstract}

\maketitle


\section{Introduction}
The Landau-Ginsburg-Wilson (LGW) theory of second order phase transitions \cite{Landau_1980,Wilson_1949} has been highly effective in describing transitions between phases across which a symmetry is broken. Although transitions of this kind are prevalent in nature, various classes of transitions that go beyond this framework have been found and studied \cite{Son_2011, Pixley_2014,Senthil_2004,Sandvik_2010,Wang_2017,Jian_2018,Qin_2017,Xu_2018,Zhang_2025}. In particular, transitions between different Symmetry Protected Topological (SPT) phases are not directly described by the LGW theory due to the lack of any symmetry being spontaneously broken across the transition and an order parameter field whose fluctuations drive the transition. A robust body of work have nonetheless used the transitions between SPT phases to search for exotic criticality and new universality classes \cite{He_2016,Grover_2013,Lu_2014,Verresen_2017,You_2016a,Morampudi_2014,Tsui_2015,Dupont_2021,Dupont_2021b,Scaffidi_2017,You_2016b,Xu_2018}.

The transition between free fermion SPT phases is relatively better understood in terms of the band structure of the fermions. A change in the band topology is accompanied by the closing of the band gap usually through a Dirac point at the transition. The Dirac point can be thought of as a delocalized version of the edge modes of the SPT phase which extend into the bulk and drive the transition. The same picture can be applied to strongly interacting SPT phases to formulate theories of their transitions\cite{Chen_2013}. Another approach is to view the critical theory at the transition as the boundary of a higher dimensional SPT state\cite{Tsui_2015b,Tsui_2019,Ji_2023}. This can be done for a large class of trivial to non-trivial SPT phase transitions in general dimensions with the bulk theory containing domain walls decorated with the non-trivial SPT phase in question. When the transition is described by a conformal field theory, this approach also relates the conformal spectrum to the topological data of the bulk. On the numerical side, a class of models known as the pivot Hamiltonians was proposed in Ref.~\cite{Tantivasadakarn_2023}. These have an enlarged symmetry at the critical point enabling a second order transition while evading the sign problem that is generically present in these transitions. 

\begin{figure*}[t]
\begin{center}
\includegraphics[width= \textwidth]{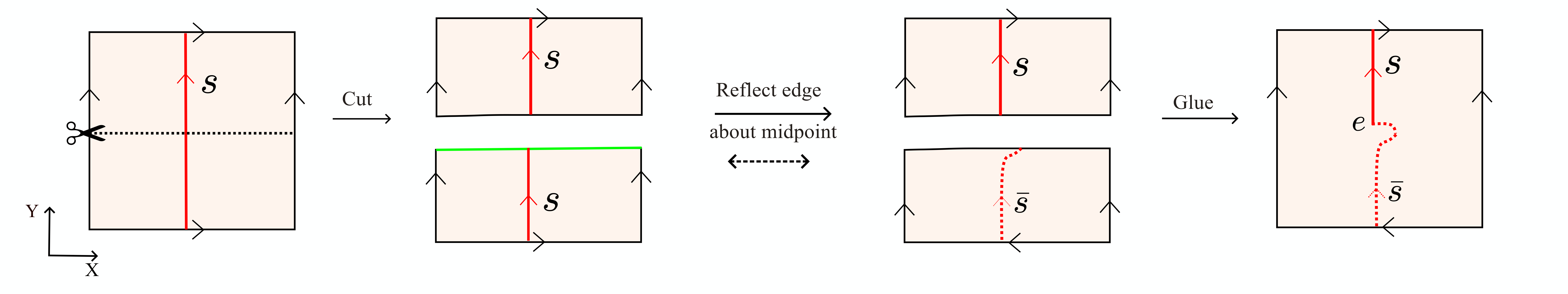}
\caption{Changing from a torus to a Klein bottle. A semion $s$ starting at the dot and tunneling across the vertical loop becomes an anti-semion $\bar{s}$ and fuses with $s$ to leave behind a charge $e$.} 
\label{fig:DSt}
\end{center}
\end{figure*}

In this paper, we focus on the kinematic aspects of the SPT phase transition associated with the symmetry response of the system rather than the dynamical aspects of it (the order of the transition, the critical exponent, etc). For LGW transitions, the kinematic data in question is simply the group of symmetries which gets broken; the connection to dynamics is provided by the LGW prescription. For SPT phase transitions, the relevant kinematic data is more than just the symmetry group. It is hidden in the spectrum of the neighboring gapped phases in the form of gapped topological defects of various codimensions. The objective of this work is to provide an illustration of how these topological defects and their corresponding invariants can place exact constraints on the universal part of the spectrum at criticality. In particular, we identify symmetry-associated topological invariants at the transition between the trivial and nontrivial $Z_2$ SPT phases in $2+1$D when the system is put on a Klein bottle. 

Our approach to studying the $Z_2$ SPT transition (and related transitions) is closely related to the notion of symmetry-enriched criticality developed in Ref.~\cite{Verresen_2021}. Since there are no local operators which can distinguish SPT phases, the authors consider non-local symmetry flux operators. The symmetry charges of these operators are topological invariants that can distinguish different SPT phases. A key insight of the authors was that these charges remain well defined at critical points. Hence this allows one to distinguish transitions with identical spectrum but whose operators carry distinct symmetry charges, for example, when two different SPT phases undergo transitions into the same symmetry broken phase. Furthermore, for certain direct SPT-SPT transitions, this implies the coexistence of distinct symmetry fluxes with identical scaling dimensions but different symmetry charges. The invariants considered in Ref.~\cite{Verresen_2021} apply when symmetry fluxes of one symmetry generator being charged under other symmetry generators and are nontrivial for example for SPT phases in $1+1$D with $Z_2^2$ symmetry and in $2+1$D with $Z_2^3$ symmetry. 

The analysis in Ref.~\cite{Verresen_2021} misses some interesting cases like $Z_2$ and $Z_2\times Z_2$ SPTs in $2+1$D. To get around this difficulty, we consider symmetry fluxes of emergent symmetries. In particular, we consider the $Z_2$ parity symmetry which emerges in the non-trivial $Z_2$ SPT phase in $2+1$D. We calculate topological invariant induced by the mixture of parity defect and $Z_2$ symmetry defect and find that it can distinguish the SPT phases. Furthermore, we also numerically verify that the invariant is well defined at criticality by detecting its signature in the form of exactly degeneracy in the ground states independent of system size. Our analysis focuses on the $Z_2$ SPT phase on the Klein bottle, but this response holds on more general closed manifolds containing orientation reversing nontrivial cycles as long as it is possible to put the SPT phase on them consistently.

The paper is organized as follows: In section~\ref{sec:SPT} we calculate explicitly the topological invariant induced by the parity defect and the $Z_2$ symmetry defect in the non-trivial $2+1$D $Z_2$ SPT phase in a concrete lattice model. In section~\ref{sec:transition}, the transition between the trivial and non-trivial SPT phases is studied and the signature of the invariant at the transition is numerically verified. In section~\ref{sec:Boundary}, we relate the symmetry response of the $2+1$D SPT state to the modular transformation of the $3+1$D $Z_2$ gauge theory using the Symmetry Topological Field Theory\cite{Ji_2020,Moradi_2022,Huang_2023,Kong_2015,Chatterjee_2023,Lichtman_2021,Lin_2023,bhardwaj_2024} framework. This extends similar connection previously established in one-lower dimension and on orientable manifolds. In section~\ref{sec:general}, we discuss the generalization of this phenomenon to other $2+1$D SPT phases. Finally in section~\ref{sec:symmetries}, we explore deeper the notion of emergent symmetry inspired by the role of parity symmetry in the $Z_2$ SPT example.

\section{Topological invariant in $2+1$D $Z_2$ SPT model}
\label{sec:SPT}

\subsection{General picture}
\label{sec:gen}

Let us start with some intuition for why a Klein bottle is useful in inducing certain symmetry response in the $2+1$D $Z_2$ SPT phase. 

A Klein bottle, as shown in Fig.~\ref{fig:DSt}, can be obtained from the 2D torus by cutting the torus open along a nontrivial cycle, rotating the boundary at one side of the cut by $\pi$ and gluing back. This process can be interpreted as inserting a defect of the parity symmetry at the location of the cut such that all excitations passing through the defect undergo orientation reversal. 

The property of the $Z_2$ SPT phase on the Klein bottle is most easily understood when the model is coupled to a $Z_2$ gauge field and turns into the twisted $Z_2$ gauge theory in $2+1$D -- the double semion topological model. The double semion topological model contains four types of anyons: the trivial one, a boson $e$, a semion $s$ with topological spin $i$ and an anti-semion $s'$ with topological spin $-i$. The boson corresponds to the $Z_2$ symmetry charge in the SPT phase. The semion and the anti-semion are flux excitations of the $Z_2$ gauge theory. That is, they correspond to end points of a symmetry defect line in the SPT phase. The fusion of the semion $s$ with the boson $e$ gives the anti-semion $s'$. Now imagine creating a pair of flux excitations $s$ in the $Z_2$ gauge theory and bringing one around a nontrivial cycle to be annhilated with the other half of the pair. If the nontrivial cycle is the orientation reserving one on the Klein bottle, as shown in Fig.~\ref{fig:DSt}, the semion $s$ excitation is mapped to the anti-semion $s'$ excitation along the cycle and fuses with the other semion not into the trivial excitation but rather into $e$. Due to this extra excitation, on the Klein bottle, the double semion ground state is only two-fold degenerate while the Toric Code (the untwisted $Z_2$ gauge theory) ground state is four-fold degenerate\cite{Chan_2016}. Translated back into the SPT picture, inserting a symmetry defect line along the orientation-reserving cycle of the Klein bottle induces an extra symmetry charge in the ground state of the system. 

We will present a more concrete calculation of this charge in the next sections. Note that the orientation-reversing loop, which is not present in the torus, is essential for the presence of this charge. Also note that this doesn't happen in the trivial SPT phase: the ground state is always charge neutral in all flux sectors on different manifolds.  

\subsection{Lattice model}
\label{sec:lattice}
Our analysis is based on the lattice model defined in Ref.~\onlinecite{Tantivasadakarn_2023} which realizes the nontrivial SPT phase with $Z_2$ symmetry in 2D. We will use this model to calculate the ground state symmetry charge in the SPT phase in this section and to study the phase transition in the next. 

\begin{figure}[ht]
    \centering
    \includegraphics[width= \textwidth]{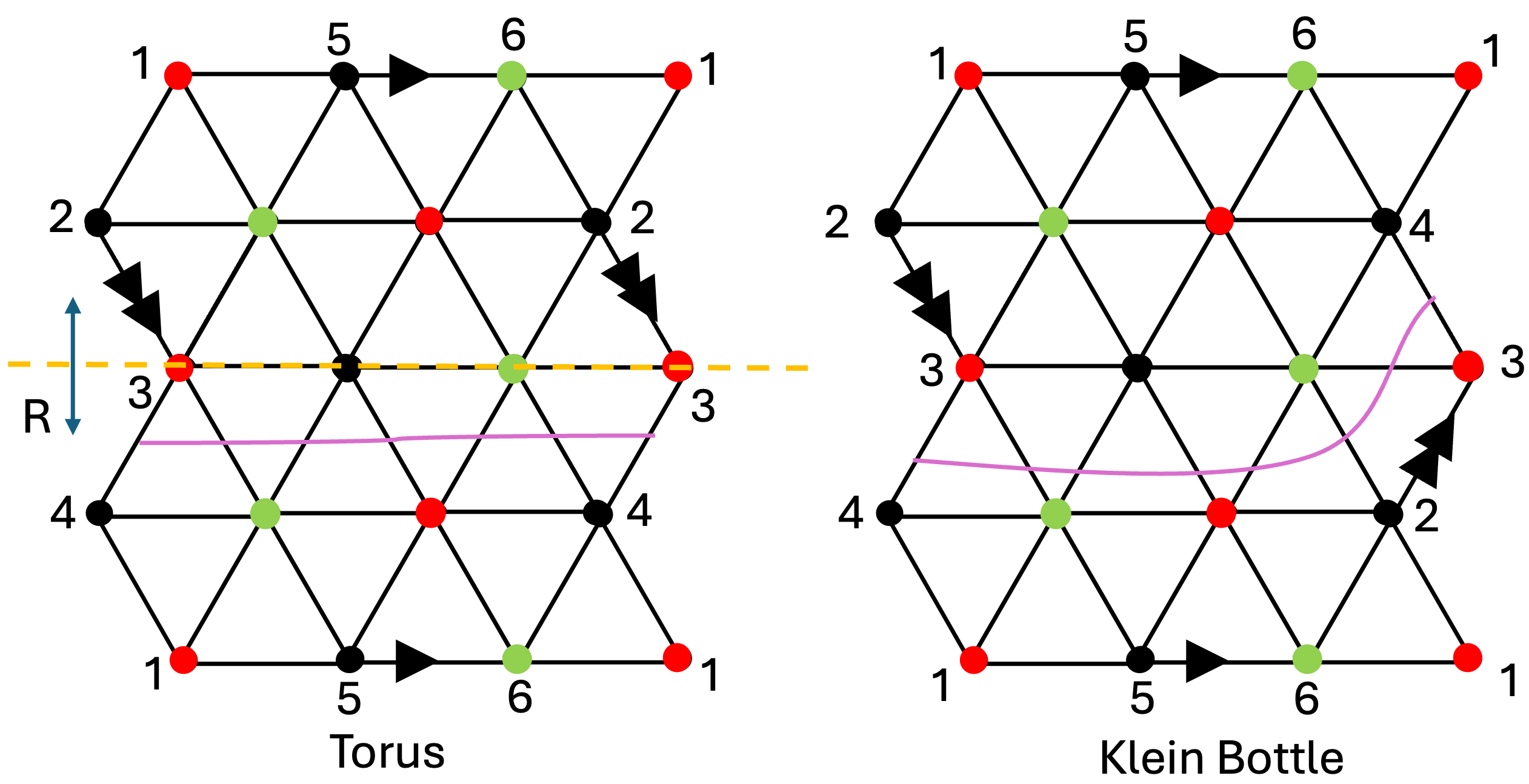}
    \caption{A tripartite (red, green, black) triangular lattice with Torus and Klein bottle boundary conditions. The arrows and the numbering of lattice sites indicate how the boundary degrees of freedom are identified. A symmetry defect can be inserted along the nontrivial purple cycle. On the Klein bottle, the defect line runs along the orientation reserving nontrivial cycle.}
    \label{fig:lattice}
\end{figure}

The model is defined on a triangular lattice with spin $1/2$s on the vertices, as shown in Fig.~\ref{fig:lattice}. The global $Z_2$ Ising symmetry is:
\begin{equation}
    S = \prod_{v} X_v 
\end{equation}
where the product runs over all vertices.
The SPT Hamiltonian is given by \cite{Tantivasadakarn_2023}
\begin{equation}
H_{\text{SPT}} = -\sum_v B_v =  -\sum_v \plaq
\label{eq:HSPT}
\end{equation}
where the red lines represent $CZ = \text{diag}(1,1,1,-1)$ gates between the adjacent vertices around the hexagon and the $X$ acts on the center vertex. It can be verified that the Hamiltonian terms commute amongst themselves and with the global symmetry using the commutation relation $CZ_{ij}X_i = X_iZ_jCZ_{ij}$. In Ref.~\cite{Tantivasadakarn_2023}, the model was viewed as an SPT with $Z_2\times Z_2\times Z_2$ symmetry. The triangular lattice is tripartite (colored red, green and black in Fig.~\ref{fig:lattice}) and the above Hamiltonian terms are actually invariant under $\prod_v X_v$ over vertices in each sub-lattice. In this work, only the diagonal $Z_2$ symmetry which acts on the entire lattice is relevant to our discussion. This particular Hamiltonian for the $Z_2$ SPT order is chosen as it displays a continuous SPT-SPT phase transition point as discussed in the next section.

Based on this lattice model, we will calculate the symmetry charge of its ground state on both the 2D torus and the Klein bottle with different flux configurations. We find that the ground state is not charged unless the system is put on a Klein bottle with nontrivial flux along the orientation-reversing cycle. We emphasize that the ground state symmetry charge is a well defined invariant of the SPT phase as the system has a unique ground state, whether on torus or Klein bottle. Hence the calculation result should be independent of the specific Hamiltonian, the lattice or the location of the symmetry defect line.

Let us start by computing the symmetry charge without flux on the torus and the Klein bottle. The left panel of Fig.~\ref{fig:lattice} shows a triangular lattice with torus boundary condition. The lattice is tri-partite, which makes the calculation of the symmetry charge of the ground state straight-forward. On a tri-partite lattice, the symmetry charge $S = \prod_v X_v$ happens to be equal to the product of all Hamiltonian terms $\prod_v B_v$. To see this, we split the latter product into three products, each over Hamiltonian terms centered on one sub-lattice. The terms in each product overlap on the red lines and the $CZ$ gates from neighboring terms cancel out, leaving $\prod_v X_v$ on each sublattice. Hence the total product is equal to the symmetry charge. Since all Hamiltonian terms are satisfied ($B_v = 1$) in the ground state, this guarantees $S = 1$. A $Z_2$ charged state $(S\neq1)$ must have at least one Hamiltonian term violated and hence cannot be the ground state. The tripartite-ness of the triangular lattice still holds on a Klein bottle, as shown in the right panel of Fig.~\ref{fig:lattice}. Therefore, the same conclusion about ground state symmetry charge applies.

\begin{figure}
    \centering
    \includegraphics[width= 0.9\textwidth]{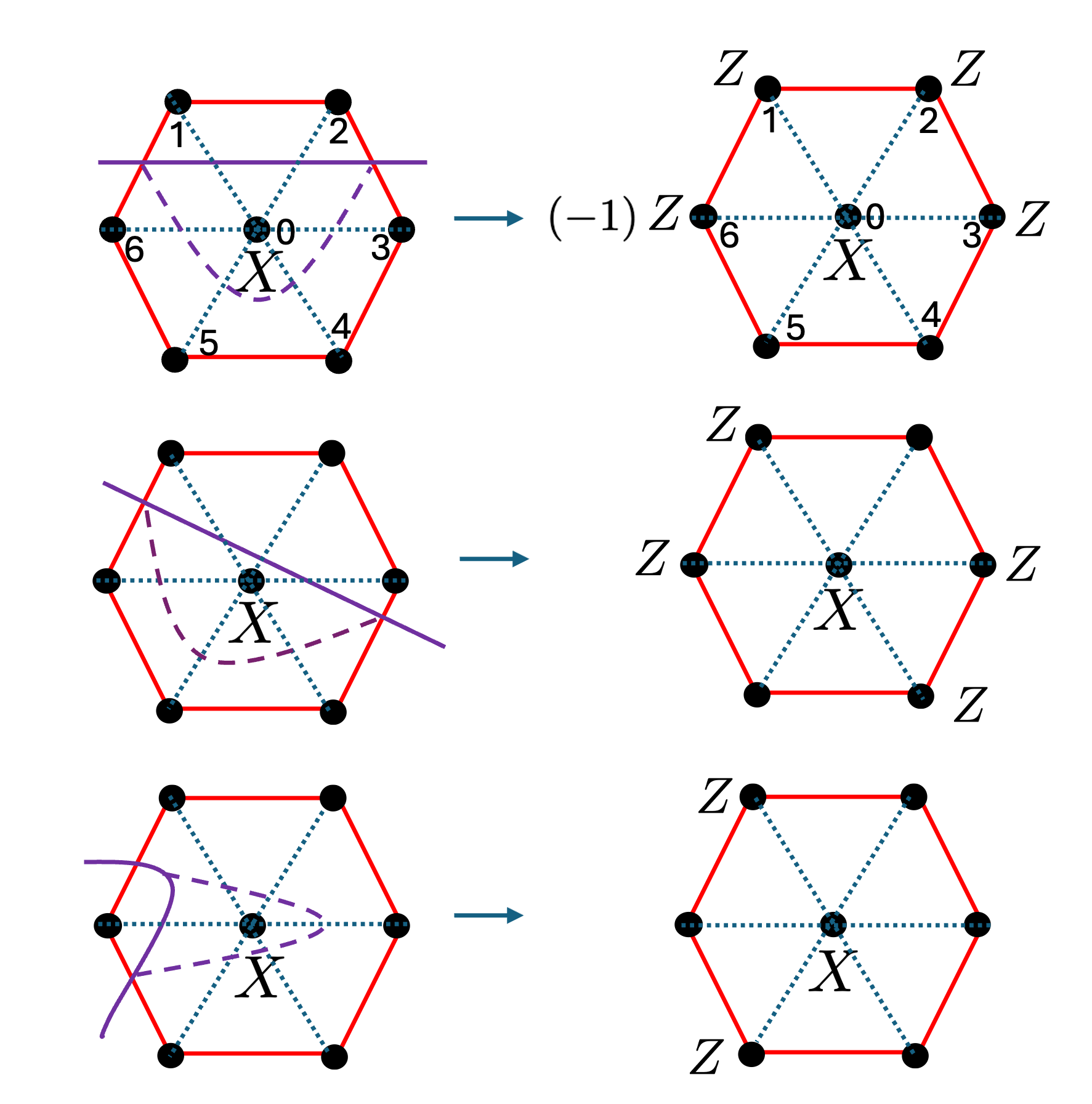}
    \caption{Action of the $\pi$ symmetry defect line (purple) on the Hamiltonian terms in $H_{\text{SPT}}$ (with the $CZ$ gates shown in red). The action is given by conjugation by the symmetry operator $\prod_v X_v$ on the vertices to one side of the symmetry defect line. The solid and dotted lines are two different symmetry defect line configurations that yield the same result. Pairs of $Z$'s are attached to the red edges crossed by the symmetry defect line. Only the first configuration, which has an odd number of triangles intersected by the symmetry defect line, picks up a negative sign.}
    \label{fig:fluxing}
\end{figure}

The next step is to thread a $\pi$ flux along a non-contractible loop. The symmetry defect line modifies the Hamiltonian terms that it intersects by conjugating one side of them by the global symmetry. Fig.~\ref{fig:fluxing} shows the distinct ways the symmetry defect line intersects a Hamiltonian term and the corresponding modifications to the Hamiltonian term. For example, the first term is modified through the conjugation of $X_1X_2$. 
\begin{equation}
\begin{array}{l}
B_v' = X_1X_2 B_v X_1X_2 \\
= (X_0CZ_{34}CZ_{45}CZ_{56})(X_1X_2 CZ_{61}CZ_{12}CZ_{23}X_1X_2) \\
= - Z_6Z_1Z_2Z_3 B_v
\end{array}
\end{equation}
The modification to the other terms can be similarly calculated and are shown in Fig.~\ref{fig:fluxing}. In general, we see that the symmetry defect line attaches pairs of $Z$'s to the red edges it crosses. Moreover, there is a minus sign when the symmetry defect line intersects an odd number of triangles. The modified Hamiltonian terms still commute with each other and the ground state still satisfies $B_v'=1$ for all $v$.

Now let's calculate $\prod_v B_v'$ and see how its relation to $\prod_v X_v$ changes. The $Z$'s attached $B_v$'s cancel each other in the product since each $Z$ show up twice along the symmetry defect line. The only difference from the previous calculation is the minus sign that occurs in certain fluxed Hamiltonian terms. Hence, $\prod_v B'_v = \pm \prod_v X_v$ depending on whether the total number of such Hamiltonian terms is even or odd. The parity of the number of these terms is related to the parity of total number of triangles intersected by the flux loop. If the total parity is odd, there must be an odd number of terms with an odd number of triangles intersected. Similarly if the parity is even, there are an even number of such terms. Hence this parity determines the ground state symmetry charge. 

Let us consider a 2D presentation of the lattice with a flux loop, such as the ones shown in Fig.~\ref{fig:lattice}. In the bulk of the lattice, the dual lattice is bipartite. That is, we can assign opposite orientation of `up' and `down' to neighboring triangles. The flux loop passes from up to down triangles and vice versa in the bulk of the lattice. The torus boundary condition respects the bipartite-ness of the dual lattice. Therefore, a closed flux loop always intersects an even number of triangles and $\prod_v B'_v = \prod_v X_v$. On a Klein bottle, the orientation reversal along one of the non-contractible loops (horizontal direction in Fig.~\ref{fig:lattice}) maps an up triangle to a down triangle and vice verse across the boundary. If there is a flux loop in this direction, it has to intersect an odd number of triangles and as a result $\prod_v B'_v = -\prod_v X_v$. A flux loop in the orientation preserving direction, on the other hand, does not lead to an extra minus sign. Therefore, we see that all four flux sectors of the torus have $+1$ symmetry charge, while on the Klein bottle, two of the flux sectors with flux along the non-orientible direction have $-1$ symmetry charge.

\subsection{Calculation using the gauged theory}

\begin{figure*}[th]
    \centering
    \includegraphics[width= \textwidth]{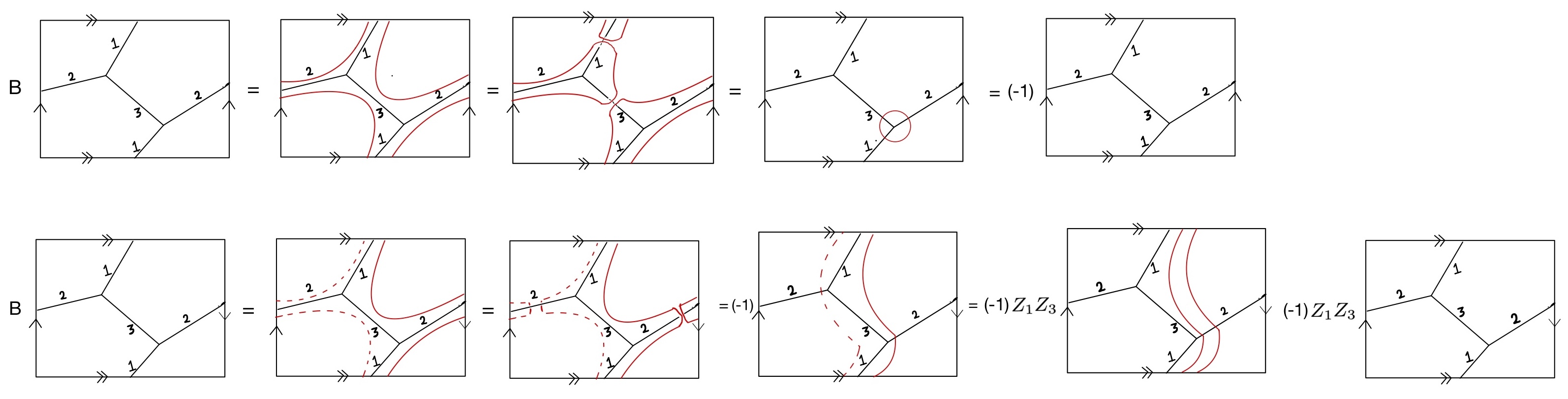}
    \caption{Action of the single plaquette term in the double semion model on a minimal lattice on the torus (top row) and the Klein bottle (bottom row). The thick (dotted) red line represent a semion string operator above (below) the lattice that is fused with it. The second equality uses the second rule in Fig.~\ref{fig:rules}. On the Klein bottle part of the semion string lies below the lattice. It is moved up by crossing the qubits on the lattice using the third rule. This results in a product of $Z$s in the fourth equality.}
    \label{fig:min}
\end{figure*}

This calculation can also be carried out in the gauged version of the SPT phase -- the double semion topological state representing the twisted $2+1$D gauge theory with gauge group $Z_2$. The symmetry charge of the SPT phase becomes the gauge charge excitation of the topological phase. The end of the symmetry defect line in the SPT phase becomes the gauge flux excitation in the topological phase. In Appendix~\ref{app:gauge}, we present a calculation based on the double semion model defined on the hexagonal lattice. In the following, we demonstrate how the same conclusion can be achieved using the model defined on the minimum lattice on the 2-torus or the Klein bottle, with two vertices and three edges\cite{Liu_2014}.

\begin{figure}[th]
    \centering
    \includegraphics[width= \textwidth]{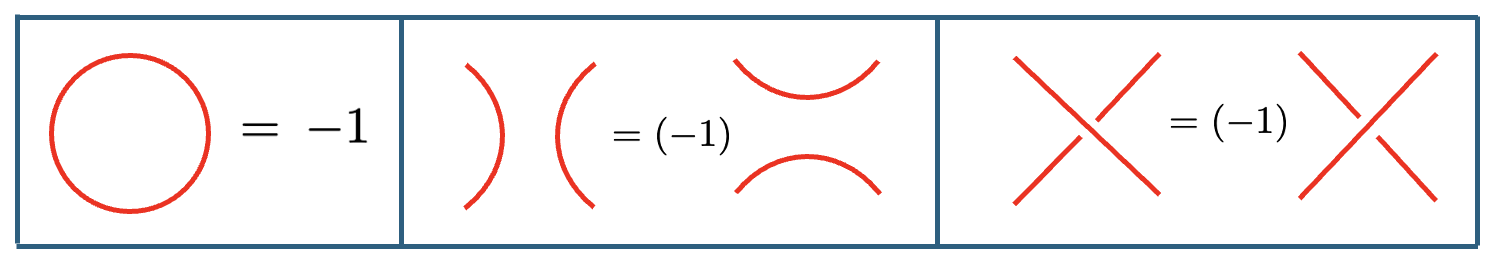}
    \caption{Rules for deforming the loop configurations in the double semion model.}
    \label{fig:rules}
\end{figure}

The double semion state defined on a tri-valent lattice contains one qubit degree of freedom on each edge. Three edges meet at each vertex in the tri-valent lattice. The $Z_2$ gauge symmetry is enforced by the vertex term
\begin{equation}
    A_v = \vertex = 1
\end{equation}
If the $|0\rangle$ qubit state is interpreted as no string along the edge and the $|1\rangle$ qubit state is interpreted as the presence of a string along the edge, the $A_v$ term enforces that the strings have to form closed loops. The deformation of the closed loops in the double semion state follows the rules given in Fig.~\ref{fig:rules}. The other Hamiltonian term defining the twisted $Z_2$ gauge theory is the plaquette term $B_p$ that comes from adding a loop to the plaquette. In the gauge theory, the vertex term measures the gauge flux at a vertex while the plaquette term measures the gauge charge inside a plaquette.

The minimal lattice contains three edge labeled $1,2,3$ in Fig.~\ref{fig:min} and can be given either the torus (top row in Fig.~\ref{fig:min}) or the Klein bottle boundary conditions (bottom row in Fig.~\ref{fig:min}). There is only one plaquette in the lattice. Adding a loop to the plaquette measures the total amount of $Z_2$ gauge charges in the whole state. The action on the state induced by adding a loop to the plaquette can be deduced using the graphical procedure illustrated in Fig.~\ref{fig:min}. On the torus, the action amounts to an overall phase factor of $-1$, while on the Klein bottle it reduced to $-Z_1Z_3$. $Z_1Z_3$ measures the amount of gauge flux through the $x$-direction nontrivial cycle -- the orientation reversing direction of the Klein bottle. Therefore, on the Klein bottle, when there is a $\pi$ flux through the orientation reversal cycle of the manifold, the double semion state gets an extra charge. No extra charge is induced on the 2-torus by the insertion of gauge fluxes. This matches with the result in Sec.~\ref{sec:lattice}.

\section{Transition}
\label{sec:transition}

\begin{figure*}[t]
    \centering
    \includegraphics[width=0.32\linewidth]{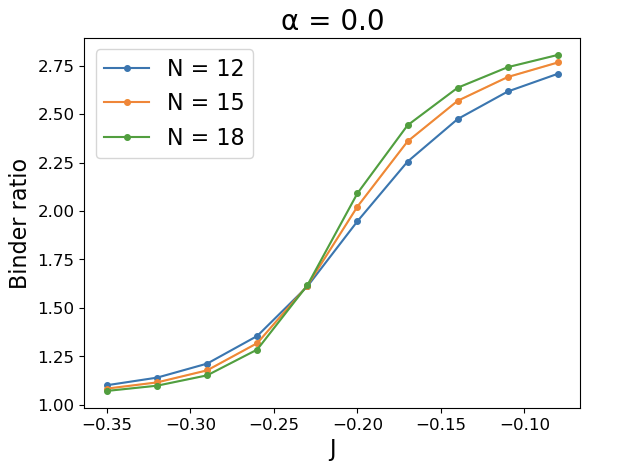}
    \includegraphics[width=0.31\linewidth]{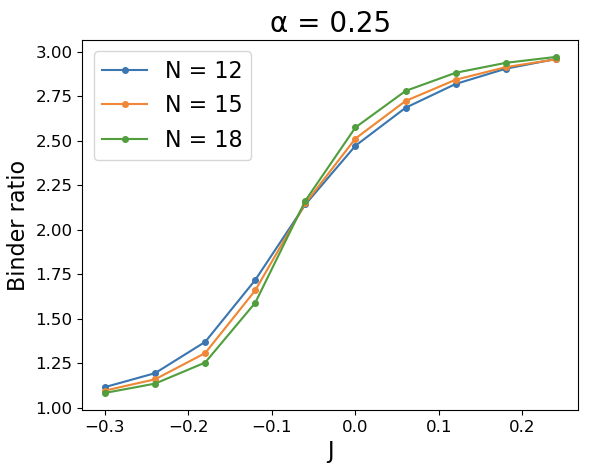}
    \includegraphics[width=0.32\linewidth]{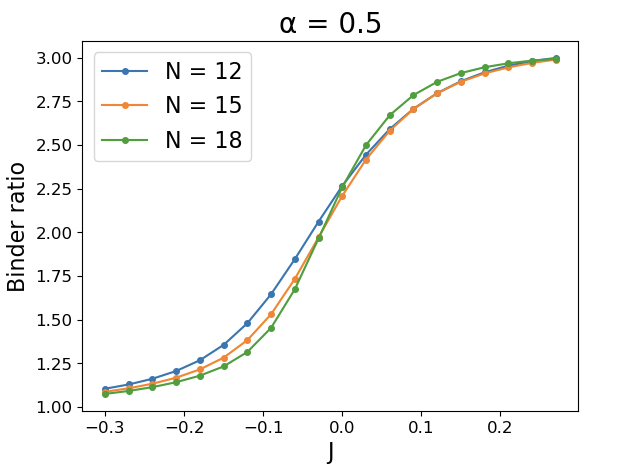}
    \caption{Binder ratio (Eq.~\ref{eq:binder}) vs J for $\alpha=0, 0.25, 0.5$. N is the total number of lattice sites. The phase boundary is determined by the intersection point of the curves.}
    \label{fig:Binder}
\end{figure*}
\begin{figure}[t]
\centering
    \includegraphics[width=0.9\linewidth]{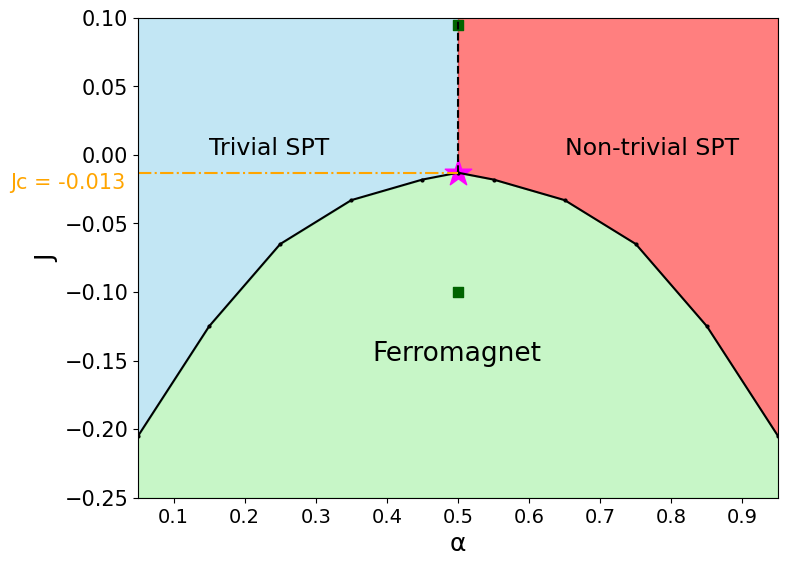}
    \caption{Phase diagram of the model in Eq.~\ref{eq:model} on a Klein bottle. The tri-critical point is marked with a star. The energy gaps at the tri-critical point as well as points marked with squares are shown in Fig.~\ref{fig:spec}}
    \label{fig:phase}
\end{figure}

In this section, the transition between the trivial and the non-trivial $Z_2$ SPT phases on the Klein bottle is numerically studied. We show that the ground-state symmetry charge discussed above leads to an exact degeneracy at the critical point. First, the entire phase diagram of a deformed model is probed using exact diagonalization \footnote{The exact diagonalization code can be found at https://doi.org/10.5281/zenodo.18241965}. The phase diagram contains the trivial and the nontrivial SPT phases as well as a ferromagnetic symmetry breaking phase. A potentially continuous transition point between the two SPT phases is identified. Then we thread a $\pi$ flux along the orientation reversing nontrivial cycle and numerically show that the trivial and non-trivial SPT phases indeed have different ground state symmetry charge. Since the $\pm 1$ charged ground states on the two sides of the SPT phase transition cannot be continuously connected, they both exist at the transition point with degenerate energy. That is, the symmetry charge protects a degeneracy at the critical point in the presence of the $\pi$ flux. Interestingly, we find that this degeneracy is exact, independent of system size. This is in contrast to, for example, the degeneracy in the ferromagnet ground space where an energy difference exists between the two symmetry breaking ground states that decays exponentially with the system size. We highlight this distinction in the numerics.

\subsection{Phase Diagram}
 
The simplest way to see a transition is by a direct interpolation between the trivial and the non-trivial SPT phases \cite{Dupont_2021b}:
\begin{equation}
    H = (1-\alpha) H_0 + \alpha H_{\text{SPT}}
\end{equation}
where \begin{equation}
 H_0 = -\sum_v X_v
\end{equation}
represents the trivial symmetric phase under the $S=\prod_v X_v$ Ising symmetry. As $\alpha$ changes from $0$ to $1$, the model transitions from the trivial to the non-trivial SPT phase. At $\alpha = 0.5$, there is an extra $Z_2$ symmetry whose effect is to swap the two SPT Hamiltonians $H_0$ and $H_{\text{SPT}}$ \cite{Bultinck_2019}. In fact, this $Z_2$ symmetry can be enhanced to $U(1)$, which plays an important role in protecting a continuous transition when other tuning parameters are added\cite{Tantivasadakarn_2023}, as shown below.

Generically, a direct interpolation of this form between SPT phases does not yield a second order transition. Indeed, in a similar lattice model for the $Z_2$ SPT phase studied in Ref.~\cite{Dupont_2021}, an intermediate gapless symmetry broken phase was observed. In order to coerce a continuous transition between the two SPT phases, nearest neighbor Ising couplings are added. \begin{equation}
    H = (1-\alpha) H_0 + \alpha H_{\text{SPT}} + J\sum_{\langle ij \rangle}Z_iZ_j
    \label{eq:model}
\end{equation}
where $\langle ij \rangle$ label nearest neighbor pairs in the triangular lattice. These terms preserve both the Ising $Z_2$ symmetry and the emergent $U(1)$ symmetry at $\alpha=0.5$. This provides an additional parameter to tune to reach an SPT-SPT phase transition point.

\begin{figure*}[th]
    \centering
    \includegraphics[width=0.4\linewidth]{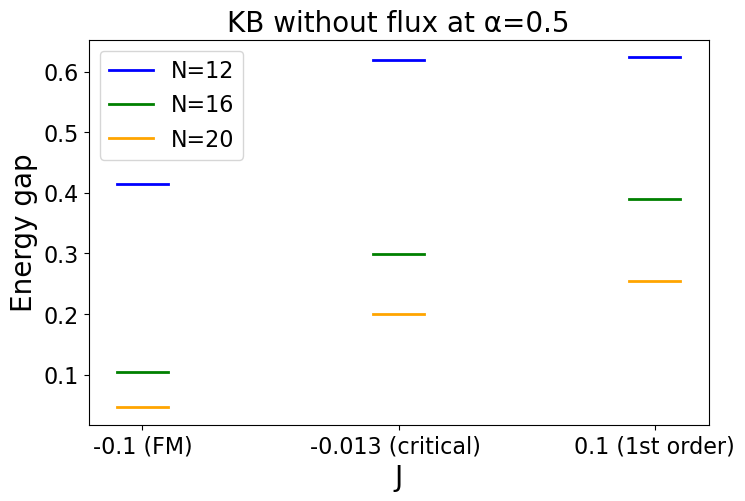}
    \includegraphics[width=0.4\linewidth]{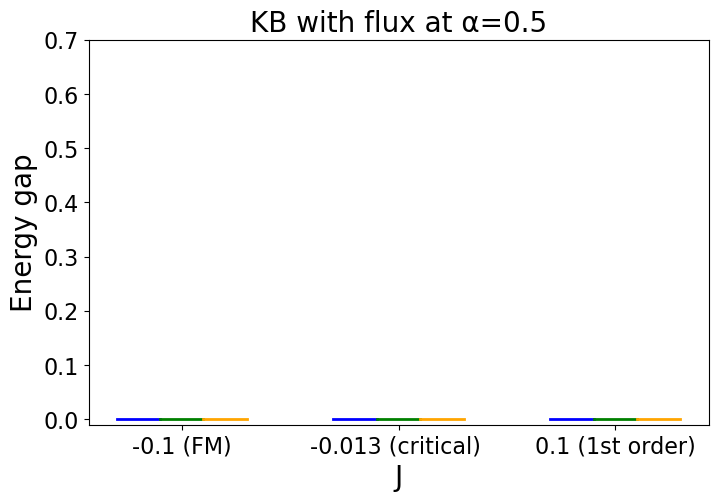}
    \includegraphics[width=0.4\linewidth]{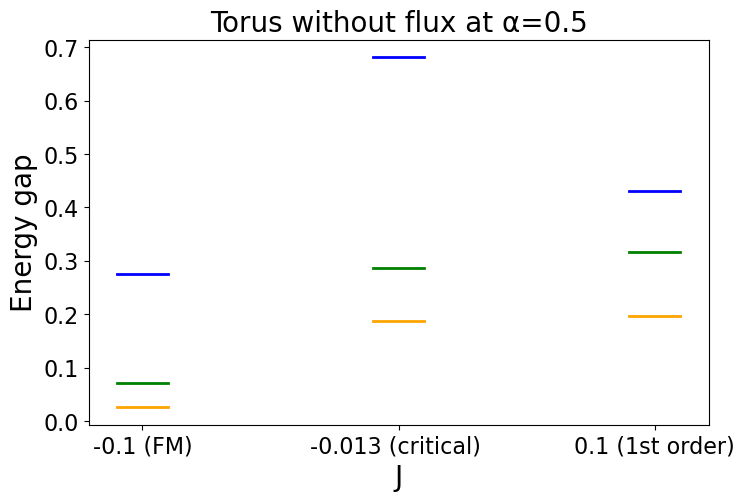}
    \includegraphics[width=0.4\linewidth]{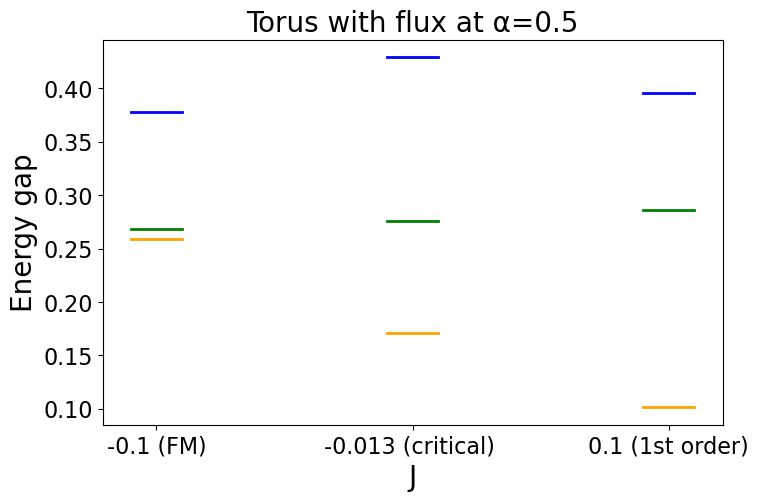}
    \caption{Energy gap for different sytem sizes along the $\alpha=0.5$ line on the Klein bottle and the torus. There is an exact degeneracy on the Klein bottle when a symmetry defect is inserted along the orientation reserving cycle. In contrast, there is no such response to the flux insertion on the torus. Indeed in all cases except KB with flux, the energy gap depends on the system size.}
    \label{fig:spec}
\end{figure*}

We now describe the phase diagram of this Hamiltonian. At $J=0$ there is a trivial SPT phase near $\alpha=0$ and a non-trivial SPT phase near $\alpha=1$. Further, due to the swap symmetry exchanging the two SPT phases, the phase boundaries of these two phases must be symmetric about $\alpha=0.5$. At a fixed value of $\alpha$ a ferromagnetic phase is expected at large negative $J$ when the nearest neighbor Ising terms dominate. This phase boundary between the SPT phases and ferromagnet phase is found numerically. Since we need access to the ground state to measure its symmetry charge, we use exact diagonalization. Due to the smaller system sizes considered, the ferromagnet order parameter is not a very accurate indicator of the transition point. We instead use the Binder ratio $B$ of the order parameter, defined as in Ref.~\cite{Binder_1981}, to locate the symmetry-breaking phase-transition line.
\begin{equation}
    B = \frac{\langle m^4\rangle }{\langle m^2\rangle^2}
    \label{eq:binder}
\end{equation}
where $\langle m^k\rangle$ is the expectation value of $(\frac{1}{L}\sum_i Z_i)^k$ on the lowest energy state obtained from the diagonalization. The crossing of the $B$ versus $J$ lines for different system sizes is an accurate way to find the transition point for a given value of $\alpha$, as shown in Fig.~\ref{fig:Binder}. The phase boundary between the SPT phases and the ferromagnetic phase found using this method is plotted in Fig.~\ref{fig:phase}. Note that, as expected from the SPT swap symmetry discussed above, the phase boundaries from the SPT phases to the ferromagnet are symmetric with respect to the $\alpha=0.5$ line. Hence the transition between the SPT phases must also occur at $\alpha=0.5$ as shown in the phase diagram. Additionally, there is a tricritical point separating the three phases at $J_c=-0.013, \alpha=0.5$.

\subsection{Critical Point}

The qualitatively similar phase diagram of this model on the torus was mapped in Ref.~\cite{Tantivasadakarn_2023}. In Ref.~\cite{Tantivasadakarn_2023}, it was found that the direct transitions between the SPT phases are all first order (shown as a dotted line $\alpha=0.5$, $J>J_c$ in our phase diagram) while the $Z_2$ symmetry breaking transitions are continuous. Additionally, the tricritical point was found to be continuous, protected by the $U(1)$ symmetry. Although we expect the same to be true in our case, we do not provide evidence for these claims. The results in this sections hold regardless of the nature of these transitions.

As discussed in \cite{Verresen_2021}, the ground state symmetry charge in flux sectors is a topological invariant at criticality. Hence the two SPT-SSB transitions are topologically distinct versions of the Ising transition. The tricritical point is at the crossover between the two criticalities and hence cannot be in the same Ising universality class. Although we do not make claims about exact universality class of the transition, the symmetry charge invariant discussed in the last section places some exact constraints on the spectrum at this critical point and the whole $\alpha=0.5$ line as we will illustrate.

First we discuss the case without flux. Along the first order crossover (the dotted line in the figure), there is typically a degeneracy in the ground state as there is a level crossing in the energy levels upon tuning from one phase to the other. For a finite size system, the levels will hybridize and there will be an energy gap exponentially small in the system size. Similarly, we expect an exponentially split degeneracy between the ground states in the ferromagnet phase. Indeed as shown in Fig.~\ref{fig:spec} the degeneracy gets more accurate with system size but the gap remains finite. At the tricritical point all the energy levels decay though at a slower rate. We also verify that the ground state has charge $+1$ throughout the phase diagram. 

Now let us thread a $\pi$ flux of the global $Z_2$ Ising symmetry along a non-orientable cycle of the Klein bottle. This amounts to modifying the Hamiltonian terms that cross the symmetry defect line according to Fig.~\ref{fig:fluxing} and running the diagonalization again. As expected, we see that at $J = J_c$ the ground state symmetry charge flips sign as we go from $\alpha < 0.5$ (trivial SPT phase) to $\alpha>0.5$ (non-trivial SPT phase). The low energy spectrum at the three points at $\alpha= 0.5$ (FM, critical and first order) are shown in Fig.~\ref{fig:spec}. Since the phases are unchanged by the addition of flux, we still expect to see a two fold degeneracy in the FM and first order points and a gapless spectrum at the critical point. However, in contrast to the no-flux sector, the two-fold degeneracy in the ferromagnetic phase and the first order transition point becomes exact and independent of system size. Moreover, this exact degeneracy shows up in the ground state of the tricritical point as well. This can be seen as a consequence of the anomaly between the Ising $Z_2$ symmetry and the SPT swap $Z_2$ symmetry. Let us label the unitary that swaps the two SPT phases as $U$ and the trivial and nontrivial SPT ground states it exchanges as $\ket{\psi_1}$ and $\ket{\psi_2}$. In the presence of the symmetry flux, since $S\ket{\psi_1}=\ket{\psi_1}$ and $S\ket{\psi_2} = -\ket{\psi_2}$, we have: 
\begin{equation}
    US\ket{\psi_1} = \ket{\psi_2}
\end{equation}
and \begin{equation}
    SU\ket{\psi_1} = -\ket{\psi_2}
\end{equation}

\begin{figure*}[th]
\centering
\includegraphics[width=\linewidth]{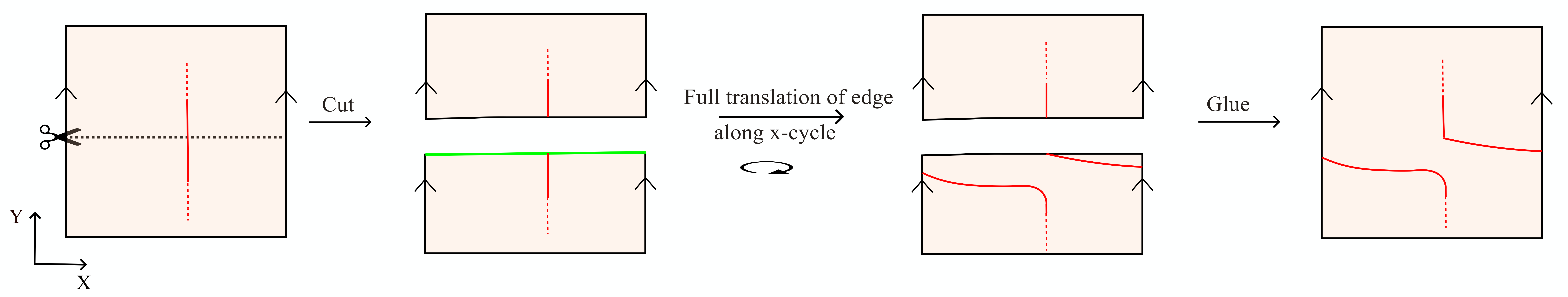}
    \caption{Dehn twist along a cylinder. The red line is shown to help visualize the translation. There are two Dehn twists on the 2-torus and only one on the Klein bottle.}
    \label{fig:dehn}
\end{figure*}

\begin{figure}[th]
\centering
\includegraphics[width=0.9\linewidth]{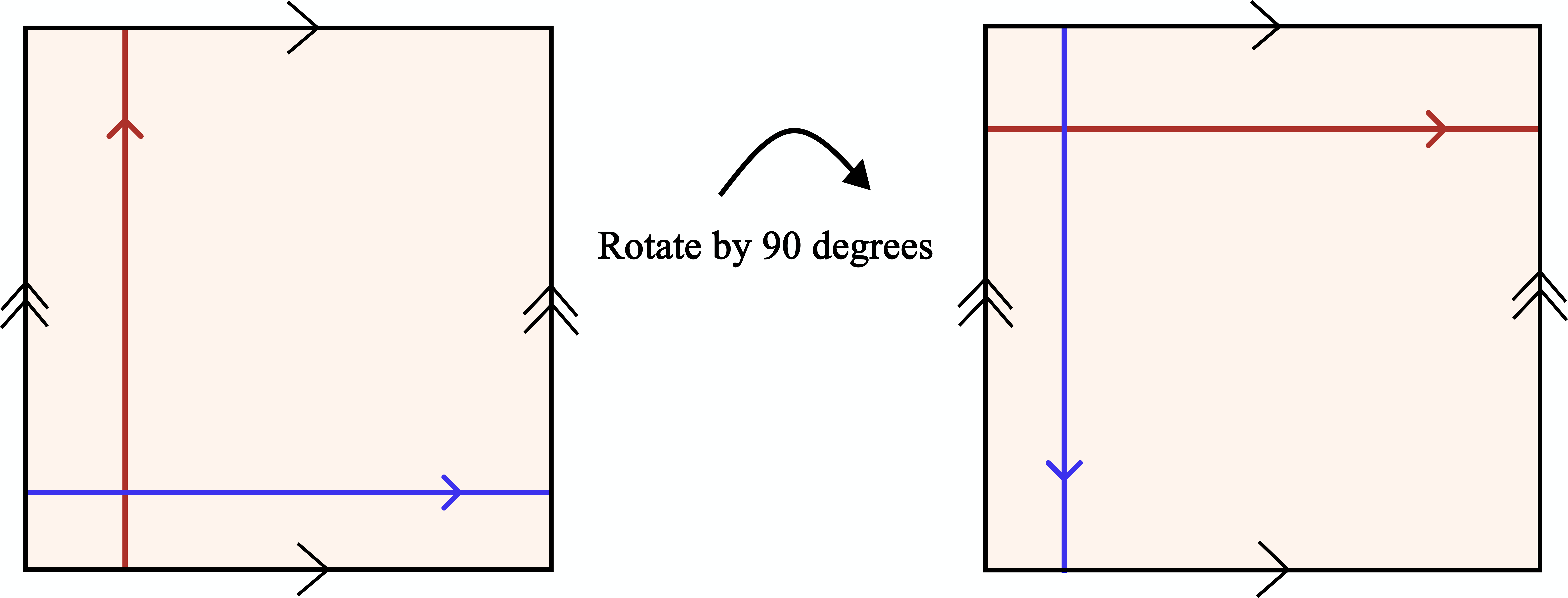}
    \caption{S transformation ($\pi/2$ rotation) on the 2-torus.}
    \label{fig:S}
\end{figure}
These states can be adiabatically evolved to $\alpha=0.5$ without crossing into the ferromagnet region. Since they carry different charges under $S$, they evolve into distinct states. At $\alpha=0.5$, the $Z_2\times Z_2$ symmetry generated by $S,U$ is represented projectively on them and hence they must be exactly degenerate. This exact degeneracy must persist as along there is a $Z_2\times Z_2$ symmetry, that is, along the entire $\alpha=0.5$ line. 

\section{Relation to modular transformations of $3+1$D Toric Code}
\label{sec:Boundary}

The symmetry charge response to the insertion of symmetry flux in the $2+1$D $Z_2$ SPT phase is closely related to the modular transformation of the $Z_2$ gauge theory in one higher dimension. We discuss in this section, how the partition function of the $2+1$D SPT phase in different $Z_2$ symmetry sectors on the Klein bottle corresponds to an invariant vector under the modular transformation of the $3+1$D $Z_2$ gauge theory on the 3D manifold of Klein bottle ($K$) $\times S^1$, following similar correspondence in lower dimensions and on orientable manifolds. First, we review in section~\ref{sec:MT_review} the correspondence in one dimension lower between the partition function of $1+1$D spin chains with $Z_2$ symmetry and the modular transformation of the $2+1$D $Z_2$ gauge theory. Next, in section~\ref{sec:MT_KxS1}, we discuss the mapping class group transformations of the 3D manifold $K\times S^1$. In section~\ref{sec:MT_SPT}, we discuss how the partition function of the $2+1$D $Z_2$ SPT phase on the Klein bottle $K$ corresponds to an invariant vector under the modular transformation of the $3+1$D $Z_2$ gauge theory on $K\times S^1$.

\subsection{Review: $1+1$D spin chain with $Z_2$ symmetry and the $2+1$D $Z_2$ gauge theory}
\label{sec:MT_review}

\begin{figure}[th]
\centering
\includegraphics[width=0.9\linewidth]{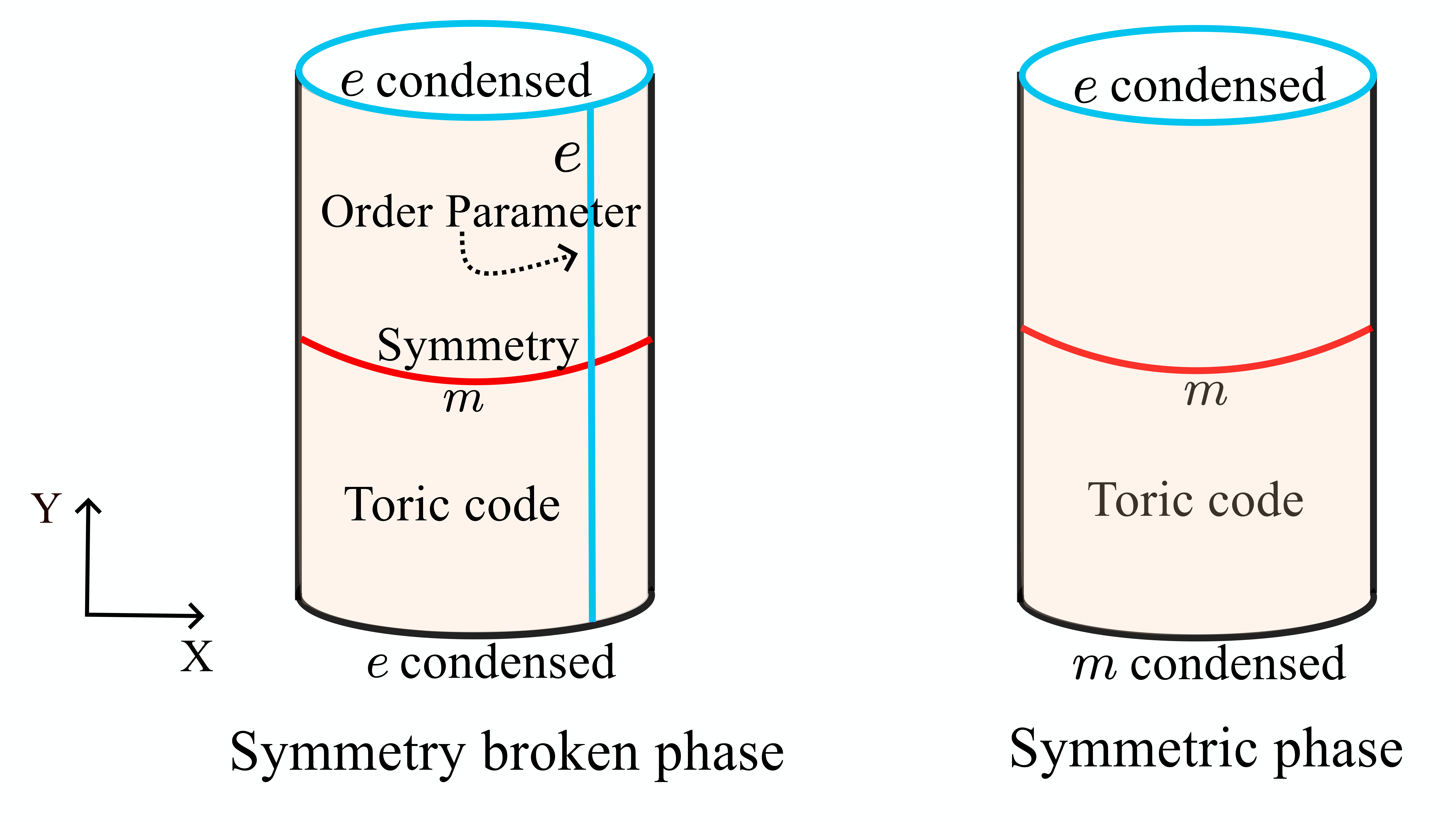}
    \caption{Gapped phases of the $Z_2$ sandwich}
    \label{fig:2Dsandwich}
\end{figure}

We start with a simple example where the correspondence is well understood: the partition function of the $1+1$D spin chain with $Z_2$ symmetry and the modular transformation of the $2+1$D $Z_2$ gauge theory. 

Consider the $2+1$D $Z_2$ gauge theory (the 2D Toric Code) defined on 2-torus ($S^1\times S^1$) in the spatial dimensions. The ground space is 4-fold degenerate and the four basis ground states can be labeled by the anyons of the topological phase $1$, $e$, $m$, $em$. The meaning of the labeling becomes clear when we choose the minimally entangled state basis \cite{Zhang_2012} of the ground space. With respect to a cut of the torus along a nontrivial cycle, for example the $x$-direction cycle, the minimally entangled states are common eigenstates of loop operators around the $x$-direction cycle. Starting from the state which has eigenvalue $1$ under all such loop operators, tunneling an anyon $\alpha$ along the other nontrivial cycle changes the eigenvalue of some of the loop operators and maps to a different minimally entangled state. Naturally, we can label the starting state as $|1\rangle$ and the resulting state as $|\alpha\rangle$, $\alpha = e,m,em$. 

The minimally entangled states provide a nice basis to write down the modular transformation on the ground space induced by the mapping class group transformations of the 2-torus. The mapping class group of the 2-torus (global transformations that keep the 2-torus invariant) has two generators: the Dehn-twist $T$ illustrated in Fig.~\ref{fig:dehn} and the $\pi/2$ rotation $S$ illustrated in Fig.~\ref{fig:S}. Together, they generate the $SL(2,Z)$ group of mapping class transformations of the 2-torus. On the ground space of the $Z_2$ gauge theory, the transformations induce unitary rotations represented by 
\begin{equation}
    S = \dfrac12\begin{pmatrix}
        1&1&1&1\\
        1&1&-1&-1\\
        1&-1&1&-1\\
        1&-1&-1&1\\
    \end{pmatrix}, T =\text{diag}(1,1,1,-1) 
\end{equation}
These modular transformation matrices are written in the minimally entangled basis with respect to an $x$-direction cut and the basis states are labeled by $1$, $e$, $m$, and $em$ respectively. The $T$ matrix is diagonal in this basis with diagonal entries corresponding to the topological spin of the anyons, while the entries in the $S$ matrix correspond to the braiding statistics between pairs of anyons.

Suppose now the 2-torus is cut open along an $x$-direction cycle into a cylinder, as shown in Fig.~\ref{fig:2Dsandwich}. It is known that the gapped boundary conditions at the top and the bottom of the cylinder are described by invariant vectors under $S$ and $T$ with non-negative integer entries and the first entry being $1$. For the $Z_2$ gauge theory considered here, there are two such vectors
$$
v_1 = (1,1,0,0)^T,\ v_2 = (1,0,1,0)^T
$$
Since the basis states are labeled by the anyons, $v_1$ corresponds to the state $|1\rangle + |e\rangle$ while $v_2$ corresponds to the state $|1\rangle + |m\rangle$. This matches with the understanding that $v_1$ represents a gapped boundary where anyon $e$ condenses while $v_2$ represents a gapped boundary where anyon $m$ condenses. The superposition $|1\rangle + |e\rangle$ represents the two sectors in an $e$ condensate: one sector with an even number of $e$, the other sector with an odd number of $e$. Similar statement holds for $|1\rangle + |m\rangle$.

In order to make connection to $1+1$D spin chains with $Z_2$ symmetry, we need to make use of the sandwich construction \cite{Ji_2020,Moradi_2022,Huang_2023,Kong_2015,Chatterjee_2023,Lichtman_2021,Lin_2023,bhardwaj_2024}. To make a $1+1$D sandwich from the $2+1$D $Z_2$ gauge theory, we put the $2+1$D theory on a cylinder with finite height, as shown in Fig.~\ref{fig:2Dsandwich}, and set the top boundary to be gapped. To discuss spin chains with $Z_2$ symmetry, we can set the top boundary to be in the $v_1$ state where $e$ is condensed. The bulk loop operator of $e$ in the $x$ direction is trivialized (set to fixed value) due to the condensation, while that of $m$ remains nontrivial (can take both $+1$ and $-1$ values). The $m$ loop operator in the $x$ direction hence acts as a $Z_2$ symmetry operator on the sandwich. The bottom boundary can then be chosen to be gapped or not gapped. Since the bottom boundary is separated from the $m$-loop in the bulk, the Hamiltonian terms near the bottom boundary, no matter how they are chosen, always respects the $Z_2$ symmetry and the sandwich represents a $1+1$D chain with $Z_2$ symmetry. 

Now if the bottom boundary is also in the $v_1$ state of $e$ condensate, the $e$ anyon can tunnel between the two boundaries without creating any excitation. Furthermore, the string operator that does the tunneling (the $y$-direction cyan line in Fig.~\ref{fig:2Dsandwich}) anti-commutes with the $x$-direction $m$ loop operator and is hence charged under the $Z_2$ symmetry. The sandwich therefore represents the symmetry breaking phase of the $Z_2$ symmetry with two degenerate ground states differing by a symmetry charge. The two components in $v_1$ at the bottom boundary, $|1\rangle$ and $|e\rangle$, correspond exactly to the two ground states in the different symmetry sectors: one in the no-charge sector, one in the charged sector. The partition function is $1$ in each of the two sectors.

\begin{figure}[th]
\centering
    \includegraphics[width=0.7\linewidth]{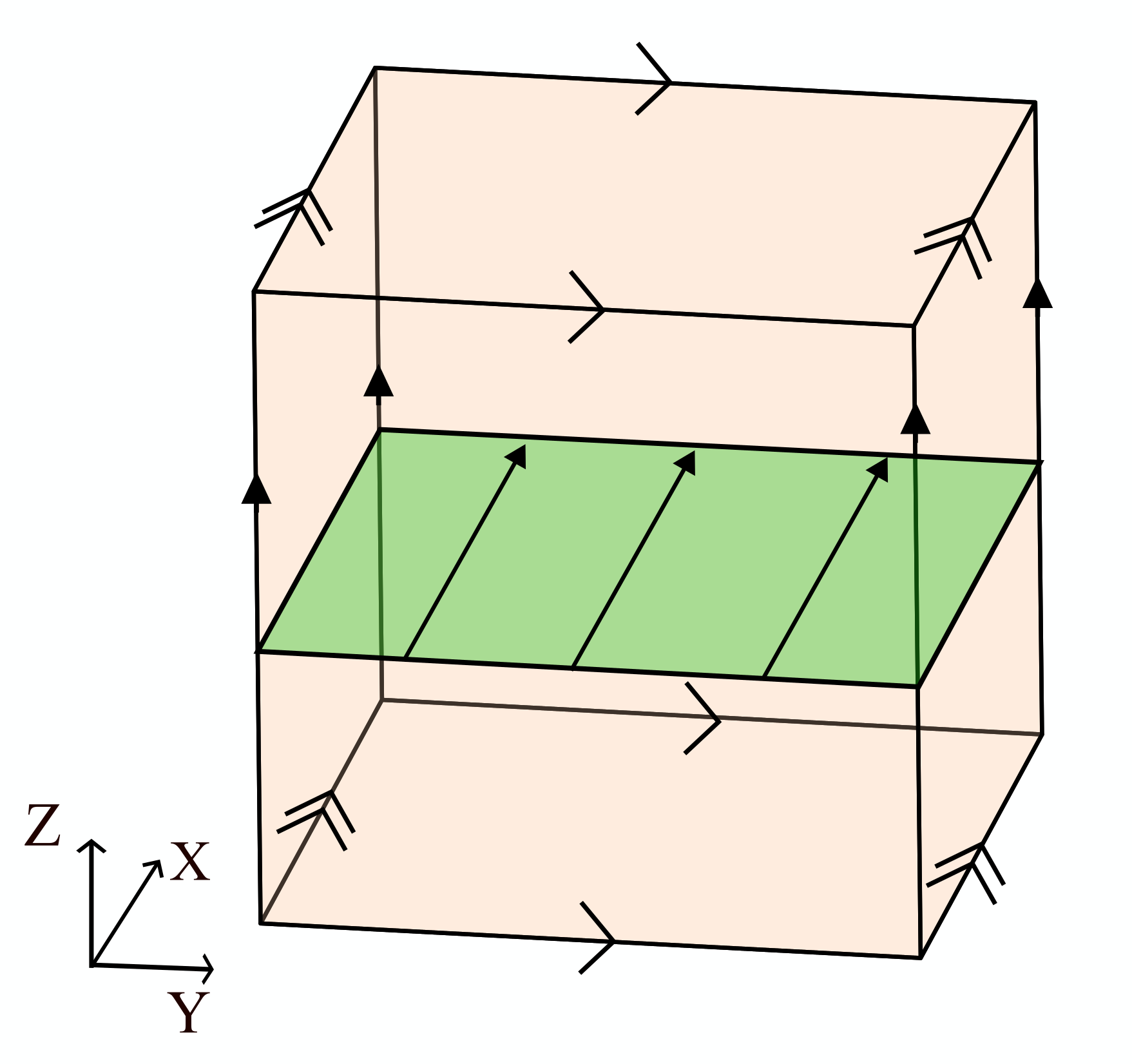}
    \caption{The green surface is a 2-torus embedded into the 3-torus. The arrows represent the direction along which a Dehn twist is performed. Here it amounts to a full translation along the $x$ cycle. It takes a non-contractible loop along $z$ to a composition of loops along both $z$ and $x$. The non-contractible loops along $x$ and $y$ are fixed.}
    \label{fig:3Ddehn}
\end{figure}

\begin{figure*}[t]
\centering
    \includegraphics[width=\linewidth]{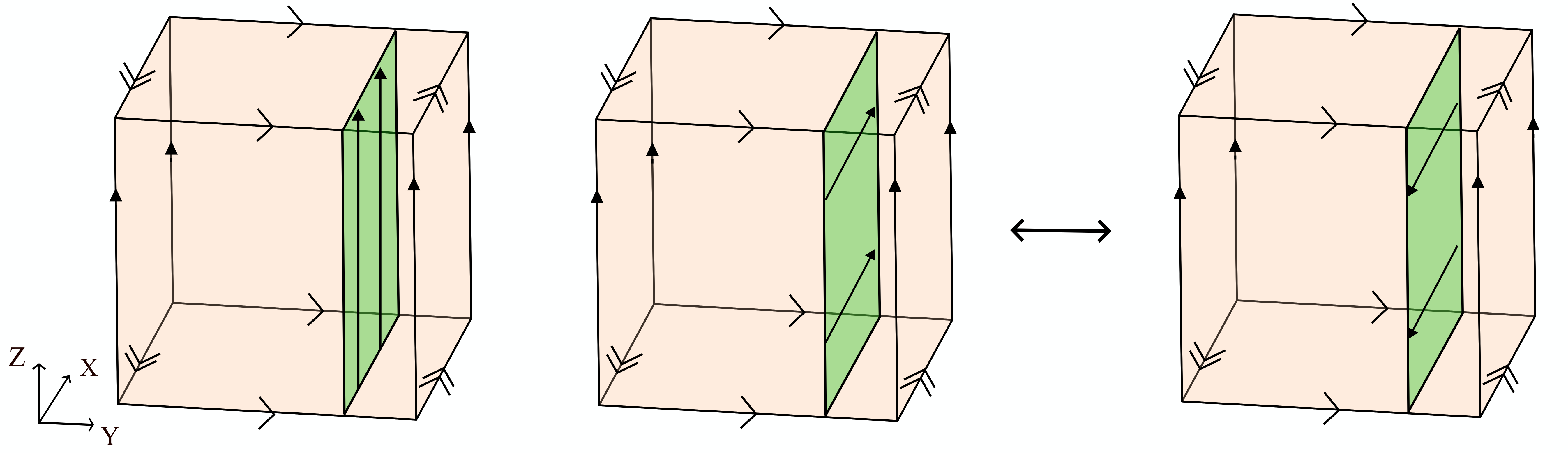}
    \caption{The green surface is an embedded essential 2-torus in $K\times S^1$. It yields two Dehn twists. The one on the right is identified with its inverse.}
    \label{fig:KBdehng}
\end{figure*}

If the bottom boundary is in the $m$-condensed state described by the $v_2$ vector, no anyon can tunneling between the top and bottom boundary. The sandwich is hence in the symmetric phase of the $Z_2$ symmetry with a unique gapped ground state. What does the $|m\rangle$ component in $v_2$ represent in terms of the partition function of the symmetric phase? Since we are using the minimally entangled basis, the $|m\rangle$ component can be obtained by tunneling an $m$ anyon in the $y$ direction starting from the $|1\rangle$ state. The tunneling of an $m$ anyon in the $y$ direction anti-commutes with the loop operator of $e$ in the $x$ direction, hence introducing a $\pi$ flux of the $Z_2$ symmetry into the $1+1$D sandwich. The $|m\rangle$ component in $v_2$ indicates that there is a unique ground state in the $\pi$ flux sector, which is what we expect for the symmetric phase of the $Z_2$ spin chain. For the symmetry breaking phase on the other hand, inserting a $\pi$ flux introduces a domain wall and takes the system out of the ground space. Therefore, there is no $|m\rangle$ component in $v_1$. 

In the next two subsections, we will see how similar correspondence can be established between the partition function of the $2+1$D $Z_2$ SPT phases and the modular transformation of the $3+1$D $Z_2$ gauge theory (3D Toric Code). The 3D manifolds of interest include both the 3-torus and $K\times S^1$. 

\subsection{Mapping class group of $K\times S^1$}
\label{sec:MT_KxS1}
\begin{figure*}[th]
\centering
    \includegraphics[width=\linewidth]{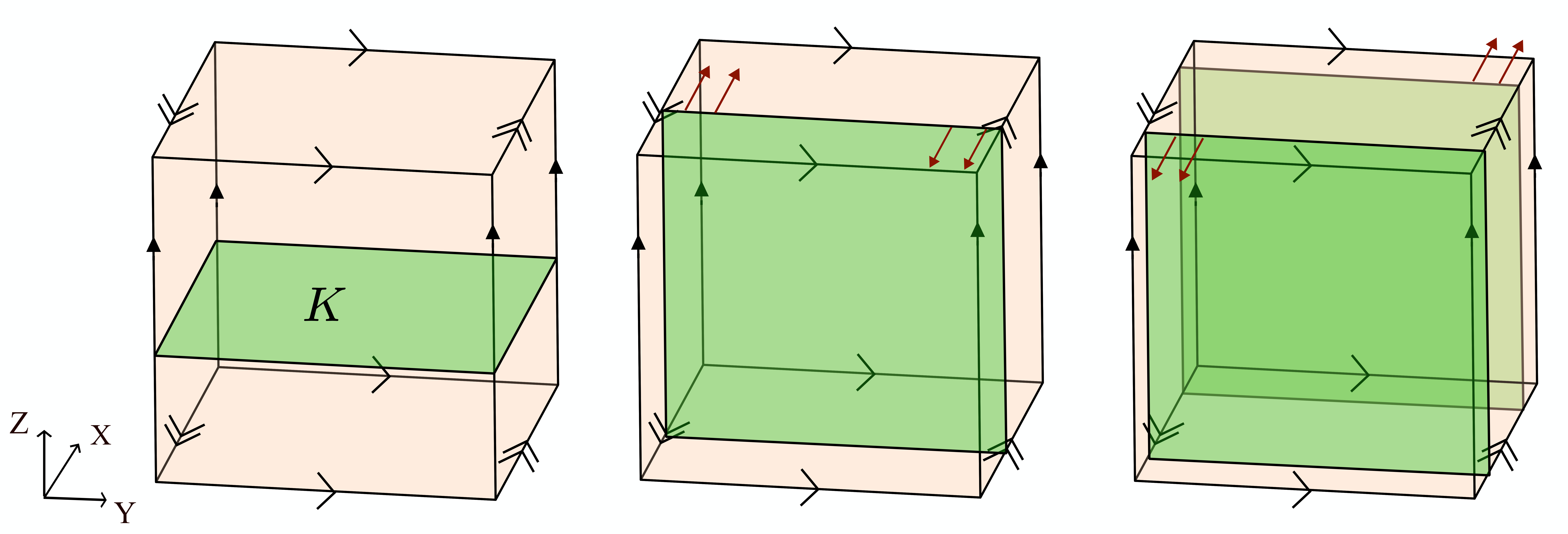}
    \caption{In $K \times S^1$, The $xy$ cross section is a Klein bottle ($K$). Two different embeddings along the $yz$ plane are shown with some normal vectors in red. The first one is non-orientable. The second one can be contracted by moving the green surfaces towards each other. Hence none of these give Dehn twists.}
    \label{fig:KBdehnb}
\end{figure*}
In order to calculate the modular transformations of the $3+1$D $Z_2$ gauge theory, we must understand the mapping class group of the 3-manifold it exists on. We first discuss the 3-torus case before moving on to $K\times S^1$. Recall that the mapping class group of the 2-torus is generated by two Dehn twists, shown in Fig.~\ref{fig:dehn}, along the two non-contractible loops. There are also two non-contractible loops on the Klein bottle but only one Dehn twist. This is because Dehn twists can only be performed along two-sided loops and one of the non-contractible loops on the Klein bottle is one-sided.

In 3D, there is an analogous result for the 3-Torus that Dehn twists generate the entire mapping class group \cite{Johannson_1979}. On $K\times S^1$ it is known that the Dehn twists generate a subgroup of the mapping class group. There is an additional order two orientation reversing transformation \cite{Paris_2014} but we disregard it as it acts trivially on the $Z_2$ gauge theory under consideration. So our task is to identify the possible Dehn twists on the 3-torus and $K\times S^1$. On the 3-Torus and $K\times S^1$, Dehn twists are performed on the so-called essential 2-tori \cite{Johannson_1979}. Essential 2-tori are 2-tori embedded in the 3-manifold which satisfy certain conditions including orientability and incompressibility, i.e. they cannot be contracted within the 3-manifold. 

We see 3 distinct ways to embed an essential 2-torus into the 3-torus, namely along the $xy, yz$ and $xz$ planes. One such embedding is shown in Fig.~\ref{fig:3Ddehn}. Dehn twists are performed by a full translation along one of the cycles of this 2-torus, shown with arrows in the figure. The twist is the identity map outside the green surface but can't be deformed to the identity on the whole manifold. There are two cycles to translate along and hence two Dehn twists for one embedding. Another way to understand this transformation is via its action on the fundamental group of the manifold. Indeed, the Dehn twist shown implements an automorphism of the fundamental group that takes a generating loop along $z$ to a composition of generating loops along $z$ and $x$. The generating loops along $x$ and $y$ are fixed. From this representation it is clear that there are at least six Dehn twists, from each choice of plane and direction of translation. These turn out to generate the full mapping class group $SL(3,Z)$. 

The situation on $K\times S^1$ is different. There is one essential torus along the $xz$ plane as shown in Fig.~\ref{fig:KBdehng}. This gives two Dehn twists with translation along $z$ and $x$. These twists act on the fundamental group generators by taking the $y$ loop to the $y$ loop plus the $z$ loop and to the $y$ loop plus the $x$ loop respectively while leaving the other generators fixed. Note that in the latter transformation, the direction of the arrows are reversed upon going around the $y$ direction. Hence the twist must be identified with its inverse and is an order two generator in the mapping class group. Along the $xy$ plane the cross section is a Klein bottle, not a 2-torus as shown in Fig.~\ref{fig:KBdehnb}. Embeddings parallel to the $yz$ plane will not be orientable as they go around the orientation reversing $y$ direction. Neither of these will give Dehn twists. Hence we expect the mapping class group to be generated by two Dehn twists along with the order two orientation reversing transformation. Indeed, by the result in Ref.~\cite{Chen_2023} the mapping class group of $K\times S^1$ is known to be $Z_2\times Z\times Z_2$.

\subsection{Partition function of the $2+1$D SPT phases}
\label{sec:MT_SPT}
\begin{figure}[t]
\centering
    \includegraphics[width=0.9\linewidth]{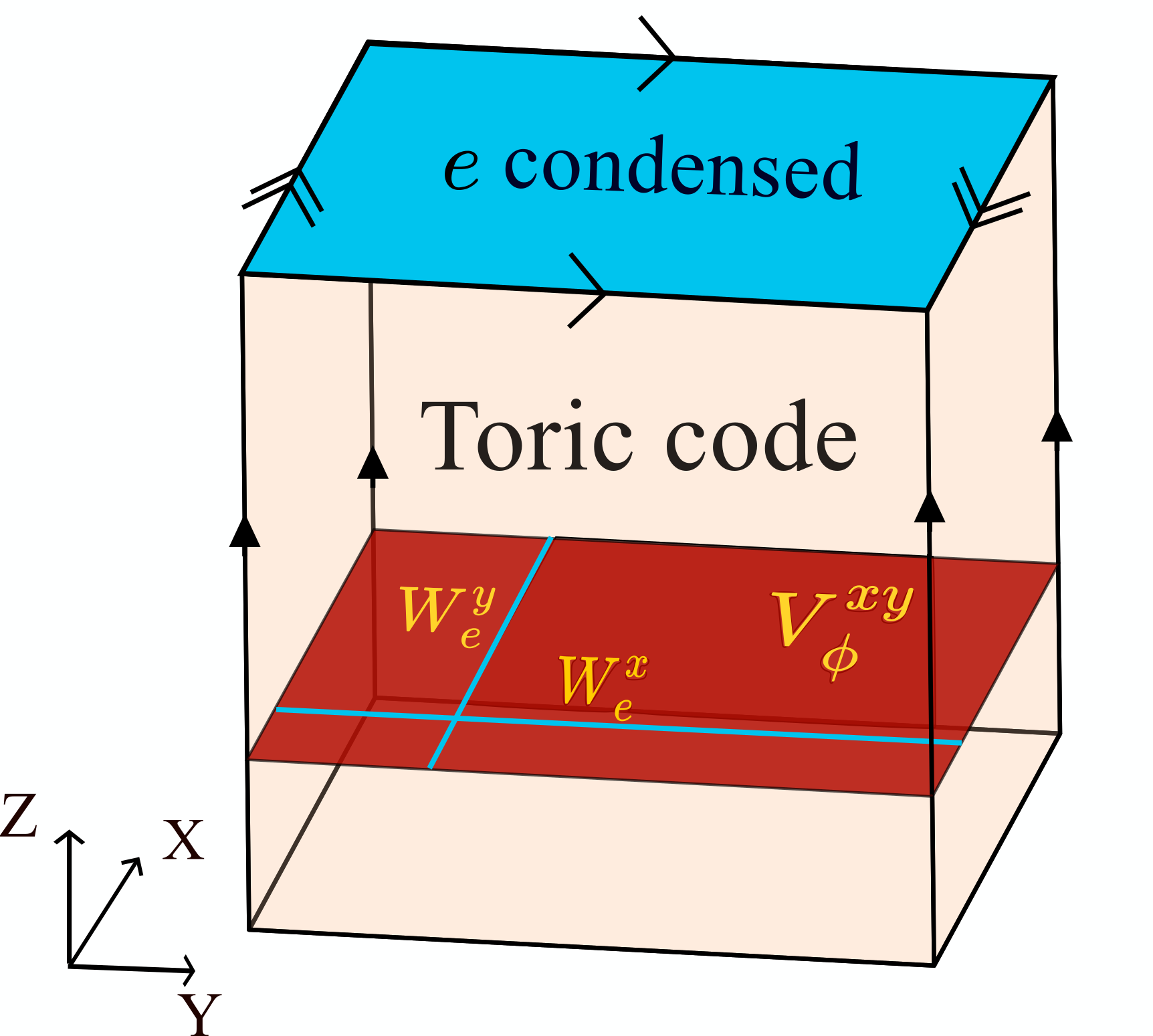}
    \caption{Sandwich for $Z_2$ symmetry in 2d. The excitations labeling the MES basis are shown.}
    \label{fig:3Dsandwich}
\end{figure}

We proceed to calculate the modular transformations of the $Z_2$ gauge theory on the 3-Torus and $K\times S^1$. This is simply the action of the mapping class groups on the finite dimensional ground space. Once we calculate these transformations, we describe how to interpret the modular invariants as partition functions of $Z_2$ symmetric phases in $2+1$D using the sandwich. 

Following Sec.~\ref{sec:MT_review}, we want to represent these transformations in the minimally entangled state basis of the ground space. Let us choose a cut perpendicular to the $z$ direction to open up the manifold into $M\times [0,1]$ where $M$ is the 2-Torus or Klein bottle. The calculation of the minimally entangled state basis is identical for both cases. This basis is labeled by the eigenvalues of the logical operators that don't cross the cut. These consist of two Wilson loops $W_e^x,W_e^y$ and a flux membrane (t'Hooft surface) $V_{\phi}^{xy}$ as shown in Fig.~\ref{fig:3Dsandwich} whose $\pm 1$ eigenvalues label the 8-dimensional ground space. For convenience let us split this space into 3 logical qubits labelled $x,y,z$. The loops $W_e^{x,y,z}$ act as Pauli matrices $Z_{x,y,z}$ and the membranes $V_{\phi}^{yz,xz,xy}$ act as $X_{x,y,z}$ respectively. Then the minimally entangled state basis is the common eigenbasis of $Z_x,Z_y,X_z$. 

Recall that in Sec.~\ref{sec:MT_review}, the minimally entangled state basis was labeled by anyons that were tunneled across the cut. A similar interpretation exists for this case. Starting from the $+1$ eigenvector of $Z_x,Z_y,X_z$ we can toggle the $Z_x$ eigenvalue by acting $X_x$, that is, by tunneling a $y$ flux loop across the cut. Similarly, the $Z_y$ eigenvalue is toggled by tunneling a $x$ flux loop. Finally, the $X_z$ eigenvalue is toggled by acting $Z_z$, that is, tunneling a charge $e$ across the cut. Hence we can also label the basis with $x,y$ flux and $e$ charge sectors.  

We focus on the 3-Torus first and calculate the modular transformations. In order to calculate how the Dehn twist act on the ground space it is convenient to momentarily work in the $Z_{x,y,z}$ eigenbasis of Wilson loops instead. Changing between the minimally entangled state basis and this one is done by acting on the last qubit by a Hadamard $H = \frac{1}{\sqrt{2}}\begin{pmatrix}
    1&1\\
    1&-1
\end{pmatrix}$. Recall from last section that each Dehn twist implements an automorphism of the fundamental group, taking one of the generating loops to a product of it with another generating loop. This translates exactly to an action on the Wilson loops; for example, the twist shown in Fig.~\ref{fig:3Ddehn} takes $W_e^z \to W_e^zW_e^x$ while fixing the other ones, $W_e^{x,y}\to W_e^{x,y}$. Additionally, the action on the $Z_2$ flux membranes is trivial. Hence on the ground space, the twist acts as a $CNOT$ gate between the $z$ and $x$ qubits, flipping the $W_e^z$ eigenvalue if the $W_e^x$ eigenvalue is $-1$. Similarly, the other Dehn twists are $CNOT$ gates between two other directions.

Now that we have calculated the bulk modular transformations we calculate their invariants, which live on the boundaries. The modular invariants on the 3-Torus are those vectors left invariant by all six $CNOT$ gates between the 3 qubits. This subspace is spanned by two vectors, which after converting to the minimally entangled state basis are listed below.

$v_e= (1,1,0,0,0,0,0,0)^T$: this vector can be interpreted as a condensate of charge ($e$) on the boundary. Indeed, it contains the vaccuum sector with even number of $e$ and the charged sector with odd number of $e$. This corresponds to the charge condensed gapped boundary, on which all the bulk Wilson lines can end.

$v_m=(1,0,1,0,1,0,1,0)^T$: this vector can be interpreted as a condensate of flux loops on the boundary. It contains all the $x,y$ flux loop sectors with no $e$ charge. This corresponds to the flux condensed gapped boundary, on which all the bulk flux membranes can end. 

Let us fix the top boundary in our sandwich to be the charge condensed boundary $v_e$. Following the discussion in Sec.~\ref{sec:MT_review}, this corresponds to $Z_2$ global symmetry in the sandwich. The flux membrane operator $V_{\phi}^{xy}$ plays the role of the symmetry operator and the loops $W_e^x$ and $W_e^y$ measure flux along $y$ and $x$ respectively. This allows an interpretation of $v_e,v_m$ as partition functions of $2+1$D $Z_2$ symmetric phases. $v_e$ contains a symmetry charged and uncharged ground state and hence is the partition function of the $Z_2$ symmetry broken phase. Indeed, in this phase charges are condensed and can be created at no energy cost while threading flux creates a domain wall with finite energy cost. $v_m$ contains ground states in all flux sectors with no symmetry charge. These are the symmetric phases, in which creating symmetry charges costs energy but threading flux does not. Hence this partition function corresponds to both the trivial and non-trivial $Z_2$ SPT phases. 

We see that the modular transformations of the theory on the 3-torus are unable to distinguish the two SPT phases. Let us repeat the calculation of modular invariants on $K\times S^1$ instead. From the previous section, we know that there are only two Dehn twists in this case and hence the modular transformations are generated by two CNOT gates: one from $z$ to $y$ and the other from $x$ to $y$. The invariant subspace is generated by five vectors. Upon converting to the minimally entangled state basis we identify seven vectors which have the form of partition functions of gapped phases.

The first two are $v_e$ and $v_m$ which once again correspond to the symmetry broken and symmetric phases. We also find the vector 
$$\tilde{v}_m = (1,0,1,0,0,1,0,1)^T$$ This is similar to the symmetric phase, but the $y$ flux sectors pickup a symmetry charge. Indeed, we recognize this charge as the signature of the non-trivial $Z_2$ SPT phase on the Klein bottle. This modular invariant also corresponds to a valid gapped boundary of the $Z_2$ gauge theory known as the twisted flux condensed boundary in the literature \cite{Ji_2023}. It is also condensate of flux loops, but differs from the flux condensed boundary in the manner of condensation. This subtle distinction is not detected by the 3-Torus modular transformations. 

Another modular invariant is vector $v_4 = (1,1,1,1,0,0,0,0)^T$. When interpreted as a partition function, it corresponds to an SPT phase with a symmetry breaking defect along a non-contractible $x$ loop \cite{Ji_2025}. Indeed, in this phase threading flux along $x$ does not cost energy and the defect ensures that both symmetry charge sectors are present. Threading flux along $y$ creates a $1$D domain wall on the defect hence costs energy. Note that this defect would explicitly break the $S$ transformation of the 2-torus hence did not show up with the 3-Torus bulk. The Klein bottle does not have any such constraint. 

Finally we have the vectors $v_5 = (1,0,1,0,0,0,0,0)^T$, $v_6=(1,1,0,0,1,0,1,0)^T$ and $v_7 = (1,1,1,1,1,1,1,1)^T$. They have no clear physical interpretations. Indeed we know that the correspondence between the invariant vectors and gapped boundaries is not one to one even in $2+1$D \cite{Kawahigashi_2015}.

\section{General SPT phases}
\label{sec:general}

The phenomenon of having exact symmetry required degeneracy at a trivial to nontrivial SPT phase transition point on a Klein bottle applies not only to the $Z_2$ SPT phase but to other 2D SPT phases as well. In this section, we identify other 2D SPT phases with abelian symmetry groups that have this phenomenon.
In order to put an SPT phase on the Klein bottle, it must admit a parity symmetry, otherwise the system cannot be gapped on the Klein bottle. This can be seen by viewing the Klein bottle as a torus with a parity symmetry defect inserted along a non-contractible loop. Excitations crossing the defect would get their orientation reversed. The defect can be gapped only if the parity reversed SPT phase is in the same phase - otherwise there will be a gapless interface separating the two sides of the defect \cite{Chan_2016}. Therefore, first we need to identify parity symmetric SPT phases. Next we must verify that threading a flux gives a charge. This is more straightforward to check after gauging the symmetry group protecting the SPT phase. In the gauged theory, we verify that there is a pure flux such that its fusion with its parity reversed counterpart is a pure charge similar to the process shown in Fig.~\ref{fig:DSt} for the $Z_2$ SPT phase. This will lead to a symmetry charged ground state in the corresponding flux sector within the nontrivial SPT phase, as well as a degeneracy at the transition to a trivial SPT phase. We expect this result to hold on more general non-orientable manifolds with symmetry flux inserted along its orientation reversing cycles as long as it is possible to put the SPT phase on them consistently.

2D SPT phases with symmetry group $G$ are classified by equivalence classes of group cocycles $[\omega] \in H^3(G,U(1))$\cite{Chen_2012}, where $[\omega]$ represents the equivalence class of cocycles containing $\omega$. An anti-unitary symmetry like parity acts on $\omega$ by complex conjugation and results in a model corresponding to $[\omega^{-1}] = [\omega]^{-1}$. For this to be in the same phase as the original model, the cocyle must satisfy $ [\omega]^{-1}= [\omega]$, that is, $[\omega]^2 = [1]$, where $[1]$ labels the trivial equivalence class of cocycles in $H^3(G,U(1))$. Hence SPT phases of order 2 within the abelian group of $H^3(G,U(1))$ admit a parity symmetry. 

For SPT phases with abelian symmetry groups, their corresponding cocycles belong to three different types. Type I cocycles involve one cyclic group $Z_N$. The possible SPT phases form a $H^3(Z_N,U(1))\cong \hat{Z}_N$ group. (We use $\hat{Z}_N$ to distinguish the group formed by the SPT phases from the symmetry group $Z_N$.) The SPT phase can be parity symmetric only if $N=2k$ is even and the phase corresponds to the $k$th element in the $\hat{Z}_N$ group. SPT phases with type II cocyles involve two groups $Z_N\times Z_M$. In these SPT phases, the fluxes of the two symmetry generators have non-trivial mutual statistics upon gauging. These statistics are trivial if the greatest common divisor of $N,M$, denoted as $(N,M)$, is equal to 1. Otherwise, such SPT phases form the group $\hat{Z}_{(N,M)}$. We will consider SPT phases which contain both type I and II cocyles. These SPT phases are classified by elements in the group $H^3(Z_N\times Z_M,U(1))\cong \hat{Z}_N\times \hat{Z}_M\times \hat{Z}_{(N,M)}$. Parity symmetric SPT phases exist for $N=2n, M=2m$. There are 8 distinct ones, characterized by the triples $(a,b,c)\in \hat{Z}_N\times \hat{Z}_M\times \hat{Z}_{(N,M)}$ where $a\in\{0,n\},b\in\{0,m\},c\in\{0,p\}$ and $p=(m,n)$. Finally, there are SPT phases corresponding to type III cocyles, which involve three groups. The form of the type III cocyles is such that flux sectors of two of the symmetry generators in orthogonal directions is charged under the third symmetry generator. The nontrivial symmetry charge response can already be detected with flux configurations on the torus as discussed in Ref.~\cite{Verresen_2021}. There is no need to make use of the Klein bottle.

To calculate how the charge and flux excitations of the gauged SPT phase get permuted by the parity symmetry, we make use of the Chern-Simons (CS) theory. All abelian topological phases can be described using the $K$ matrix formalism of the CS theory, which we briefly review here \cite{Wen_2004}. The low energy effective action of an abelian topological phase on spacetime $M$ is given by the $U(1)^N$ CS action:
\begin{equation}
    S[a] = \frac{1}{4\pi}\int_MK_{ij}a^{(i)}\wedge da^{(j)}
\end{equation} where $a^{(i)}$ is the $i$th $U(1)$ gauge field. The integer matrix $K$ is symmetric and invertible. An anyon excitation is labeled by an integer vector $l\in Z^{N}$. The braiding and self statistics data of the anyons can be extracted from the $K$ matrix and are given by: 
\begin{equation}
\label{braid}
    s_{l,l'} = 2\pi l^TK^{-1}l'
\end{equation}
\begin{equation}
    t_l = \pi l^TK^{-1}l
\end{equation}
Excitations labeled by vectors of the form $Kl$ have trivial braiding with all other anyons and hence can be identified with local (non-fractional) excitations. Excitations which differ by these particles are identified with each other, leaving behind a finite set of distinct anyon excitations. The diagonal entries of the $K$ matrix must be even integers to describe bosonic topological phases.

For a given phase, a relabeling of all the anyons $l\to W^{T}l$ combined with a transformation of the $K$ matrix $K\to W^TKW$ does not change any observable statistics for matrices $W\in GL(N,Z)$. $GL(N,Z)$ is known as the unimodular group and consists of matrices with integer entries and determinant of unit magnitude. Therefore, the phases described by $K$ and $K' = W^TKW$ are the same.

Since parity acts by complex conjugation of the self and mutual statistical phase factors, given a phase corresponding to $K$, one representative of the parity reversed phase is $-K$. Hence an abelian topological phase is parity symmetric if there is a $W\in GL(N,Z)$ such that \cite{Chan_2016}: 
\begin{equation}
\label{inv}
    W^TKW = -K
\end{equation} 

The $K$ matrix of the (gauged) parity symmetric type I SPT phase with symmetry group $Z_{2k}$ is 
\begin{equation}
    K = \begin{pmatrix}
        -2k & 2k\\
        2k & 0
    \end{pmatrix}
\end{equation}
This reproduces the self statistics of the elementary flux excitation $s = \begin{pmatrix}
    0&1
\end{pmatrix}^T$:  $\theta_s = \dfrac{\pi}{2k}$ and charge-flux braiding of $\theta_{e,s} = \dfrac{\pi}{k}$ for charge $e = \begin{pmatrix}
    1&0
\end{pmatrix}^T$. Parity symmetry can be realized by $W = \begin{pmatrix}
    1 & 0 \\
    1 &-1
\end{pmatrix}$ which satisfies Eq.~\ref{inv}. Under the action of this symmetry, a pure flux $\begin{pmatrix} 0&1 \end{pmatrix}^T$ is transmuted to the dyon $\begin{pmatrix} 1&-1 \end{pmatrix}^T$. This fuses with the original flux to leave a pure charge $\begin{pmatrix} 1&0\end{pmatrix}^T$ as required.

The same $K$ matrix analysis goes through for SPT phases with both type I and type II cocycles. Following the notation above, let us consider an order 2 SPT phase $(a,b,c)\in \hat{Z}_{2n}\times \hat{Z}_{2m}\times \hat{Z}_{2(n,m)}$ with $c=(m,n)>1$ and $a\in\{0,n\},b\in\{0,m\}$. This has K matrix:
\begin{equation}
    K = \begin{pmatrix}
        -2a &2n&-c&0\\
        2n&0&0&0\\
        -c&0&-2b&2m\\
        0&0&2m&0
        \label{eq:Kabc}
    \end{pmatrix}
\end{equation} The second and fourth coordinates are the fluxes while the first and third are the charges. Since $c=(m,n)$, there exist integers $u,v$ such that $um+vn=c$. Let us define the indicator functions $i(x)=0$ if $x=0$ and $i(x)=1$ otherwise. The matrix $U = \begin{pmatrix}
    1&0&0&0\\
    i(a)&-1&u&0\\
    0&0&1&0\\
    v&0&i(b)&-1
\end{pmatrix}$ satisfies equation \ref{inv} for the $K$ matrix in Eq.~\ref{eq:Kabc} and is a possible presentation of the parity symmetry. Under its action, the flux of the first $Z_{2n}$ symmetry, given by $\begin{pmatrix}
    0&1&0&0
\end{pmatrix}^{T}$ is transmuted to $\begin{pmatrix}
    i(a)&-1&u&0
\end{pmatrix}^{T}$. Similarly, the flux of the $Z_{2m}$ symmetry,  given by $\begin{pmatrix}
    0&0&0&1
\end{pmatrix}^{T}$ is transmuted to $\begin{pmatrix}
    v&0&u\cdot i(b)&-1
\end{pmatrix}^{T}$. These fluxes fuse with their parity reversed counterpart to leave behind pure charges.

Hence the features discussed above for the $2+1$D SPT phase on the Klein bottle apply to all type I and type II SPT phases of order 2 with abelian symmetries.

\section{Emergent symmetry}
\label{sec:symmetries}

It is interesting to consider the role played by parity symmetry in this story and compare it to the role played by the $Z_2$ Ising symmetry. In this section, we introduce the concept of `defining symmetry' and `emergent symmetry' to clarify their difference. We discuss the relation between `emergent symmetry' and the concept of `low-entanglement excitation' introduced in Ref.~\cite{Stephen_2024,Ji_2025,Zhao_2025} as well as its realization in the Symmetry TFT formalism. 

\subsection{Defining vs. emergent symmetries}

The SPT phase represented by the Hamiltonian in Eq.~\ref{eq:HSPT} is protected by the $Z_2$ symmetry and the $Z_2$ symmetry only. If the $Z_2$ symmetry is broken, the SPT phase can be smoothly connected to a trivial phase (with a product state ground state). If the $Z_2$ symmetry is not broken, the SPT phase is nontrivial and cannot be smoothly connected to a trivial phase no matter how the Hamiltonian is deformed. We say that the $Z_2$ symmetry is a \textit{defining} symmetry of the SPT phase. If the $Z_2$ symmetry is broken, the phase no longer exists. 

On the other hand, the parity symmetry is not crucial for the existence of the SPT phase. The SPT Hamiltonian can be invariant under a parity transformation. For example, the Hamiltonian in Eq.~\ref{eq:HSPT} (on a 2D torus) is symmetric under up-down reflection which we denote as $R$ and the ground state $|\psi\rangle$ is invariant under $R$.
\[
R|\psi\rangle = |\psi\rangle
\]
We can require this to be an extra symmetry of the Hamiltonian, but we don't have to. If this symmetry is explicitly broken, as long as the Hamiltonian is still $Z_2$ symmetric, the SPT order still exists and there is no way to smoothly deform the phase to a trivial one. In fact, since the Hamiltonian remains in the same $Z_2$ gapped phase, the new ground state $|\psi'\rangle$ can be `quasi-adiabatically continued'\cite{Hastings_2004,Hastings_2005} to the original $|\psi\rangle$ by a $Z_2$ symmetric finite-depth quantum circuit $U$, $|\psi'\rangle = U|\psi\rangle$. If we conjugate $R$ by $U$, we obtain a different parity symmetry operator $R' = URU^{\dagger}$ such that $|\psi'\rangle$ is invariant under the new parity symmetry operator
\[
R'|\psi'\rangle = URU^{\dagger}U|\psi\rangle = |\psi'\rangle
\]
Note that since $U$ can break the original parity symmetry, $R'=URU^{\dagger}$ can be different from $R$. Apart from the up-down reflection of the lattice, it can also contain a non-spatial part in the form of a finite depth quantum circuit. What is potentially lost with $R'$ is the exact locality within the tensor product Hilbert space. That is, a single-site operator might be mapped to a few-site `fattened' operator due to the finite depth quantum circuit part of $R'$, but locality in general is still preserved. A local operator remains local under $R'$. 

If we accept that a symmetry operator does not have to preserve the exact locality of the Hilbert space, then as long as the system remains in the $Z_2$ SPT phase, it always has a parity symmetry of some form. That is, the parity symmetry is \textit{emergent}. It is very important to note that this symmetry emerges only in the low energy ground state of the SPT phase, not in the full Hilbert space. This is because in the above argument, the existence of $U$ only holds for the ground state\cite{Hastings_2004,Hastings_2005}. It is a low-energy property of the SPT phase.

The situation is very similar to how higher-form symmetries\cite{Gaiotto_2015} emerge in topological phases. Consider, for example, the 2D $Z_2$ topological order represented by the toric code model. In this case, the defining symmetry is trivial. That is, the topological phase cannot be smoothly connected to a trivial phase no matter how the Hamiltonian is deformed. On the other hand, the fixed point toric code Hamiltonian
\begin{equation}
  \begin{aligned}
H  = & -\sum_p A_p - \sum_v B_v \\
=&-\sum_p \prod_{e\in p} X_e - \sum_v \prod_{v \in e} Z_e
\end{aligned}  
\end{equation}
where $e$, $v$ and $p$ denote the edges, vertices and plaquettes in a 2D lattice, has an explicit 1-form symmetry given by all the loop operators of the form
\[
W_{\mathcal{L}} = \prod_{e\in\mathcal{L}} Z_e
\]
where $\mathcal{L}$ represents both contractible and non-contractible closed loops on the dual lattice. We can explicitly preserve this symmetry, but we don't have to. Adding local terms that violate the $W_{\mathcal{L}}$ operators does not change the topological order of the model as long as the Hamiltonian remains gapped. In fact, the same  quasi-adiabatic continuation argument shows that as long as the system remains in the same topological phase, the degenerate ground states $|\psi'_i\rangle$ are related to the fixed point ground states $|\psi_i\rangle$ through a finite depth quantum circuit $U$, $|\psi'_i\rangle = U|\psi_i\rangle$. Therefore, we can always redefine the 1-form symmetry as
\[
W'_{\mathcal{L}} = UW_{\mathcal{L}}U^{\dagger} 
\]
such that the $|\psi'_i\rangle$'s transform under $W'_{\mathcal{L}}$ in exactly the same way the $|\psi_i\rangle$'s transform under $W_{\mathcal{L}}$. We say that the 1-form symmetry is emergent in the whole topological phase.

Note that we can choose to impose an explicit lattice form of an emergent symmetry on the class of Hamiltonian we consider. In this case, the emergent symmetry becomes part of the defining symmetry and the emergent symmetry of the system is reduced.

\subsection{Characterization using low entanglement excitations}
The above discussion seems to suggest that given a gapped phase with its defining symmetry, we can find emergent symmetries by looking at a particular Hamiltonian realizing the phase and identify any transformation (apart from the defining symmetry) that keeps this particular Hamiltonian invariant. Then using quasi-adiabatic continuation, we can extend the transformation to be an emergent symmetry of the entire phase. But a particular Hamiltonian has a lot of invariance. For example, the $Z_2$ SPT Hamiltonian in Eq.~\ref{eq:HSPT} is invariant under the transformation induced by each Hamiltonian term (because all the Hamiltonian terms commute). Should we include all of them when considering emergent symmetries of the $Z_2$ SPT phase? Moreover, an emergent symmetry can be easier to spot on some Hamiltonians than others. How do we know if we have found all the emergent symmetries of a phase?

We propose that the interesting emergent symmetries are the ones that have symmetry defects / symmetry twists given by non-trivial `low-entanglement excitations' (LEEs) defined in Ref.~\cite{Stephen_2024,Ji_2025,Zhao_2025}. LEEs are 0d, 1d, 2d... excitations on top of the ground state of a gapped Hamiltonian that preserve the entanglement area law of the ground state. They are gapped ground states of a new Hamiltonian, which is obtained from the original Hamiltonian through modification along the dimension of the excitation. Non-trivial LEEs are excitations that cannot be created with a finite-depth circuit along the dimension of the excitation. They are either created at the boundary of a higher-dimensional circuit or through a sequential circuit in the original dimension. LEEs related by a finite-depth circuit in the original dimension are equivalent. The emergent symmetries are then implemented by sweeping the symmetry defect / symmetry twist given by the LEEs through space. Note that since we only consider the effect LEEs have on top of the ground state, it reflects only the low-energy property of the phase. The corresponding emergent symmetries emerge only at low energy.

Consider, for example, the symmetry breaking phases in a system with $Z_2$ Ising symmetry. The nontrivial LEE in this phase is the domain wall between spin up and spin down domains. A domain wall cannot be created locally and is hence a nontrivial LEE. The domain wall can be moved by applying the spin flip operator in neighboring domains, therefore the emergent symmetry in this case is the $Z_2$ Ising symmetry -- the same as the defining symmetry. This is a general feature for symmetry breaking phases.

Consider next the 1-form symmetry in the $Z_2$ topological phase. The LEE in this case is the point excitation of gauge charge $e$. It is a non-trivial LEE because the gauge charge $e$ cannot be generated by a local unitary transformation. Instead, it can only be generated at the end of a string operator -- a 1d finite-depth circuit. Moving the excitation around a closed loop implements the 1-form symmetry. 

The $Z_2$ topological phase also has an interesting 0-form emergent symmetry -- the $Z_2$ exchange symmetry between the gauge charge $e$ and the gauge flux $m$. The Hamiltonian in this topological phase does not need to have this symmetry. For example, the $e$ and $m$ excitations can have different energies. But in the ground state, this symmetry always emerges. As shown in Ref.~\cite{Roumpedakis_2023}, the Toric Code ground state has a nontrivial invertible 1d defect which can be generated through `higher-gauging'. As shown in Ref.~\cite{Tantivasadakarn_2024}, this defect cannot be generated with a 1d finite depth circuit. It needs to be generated with a 1d sequential circuit, hence a nontrivial 1d LEE. Sweeping this defect through the 2d space implements the exchange symmetry between $e$ and $m$. 

\subsection{Emergent parity symmetry in $Z_2$ SPT phase}

Following the previous argument, we show in this section that parity symmetry is an emergent symmetry of the 2d $Z_2$ SPT phase. In particular, we show that the symmetry defect is generated through a 1d sequential circuit. Note that this seems to contradict with the result in Ref.~\cite{Ji_2025} where one of us (together with co-authors) claimed that different SPT phases have the same LEEs. In the trivial 2d $Z_2$ SPT phase, the only nontrivial 1d LEE is a 1d $Z_2$ symmetry breaking defect. But the parity symmetry defect we consider here does not break the $Z_2$ symmetry, so it is a new kind of 1d LEE. This happens because the argument in Ref.~\cite{Ji_2025} does not apply here. The argument in Ref.~\cite{Ji_2025} makes use of the fact that the ground state in the trivial SPT phase $|\psi_0\rangle$ can be mapped to the one in the nontrivial SPT phase $|\psi_1\rangle$ through a symmetric Quantum Cellular Automaton (QCA, a locality preserving map) $U$. Therefore, if an LEE can be generated on top of $|\psi_0\rangle$ using circuit $V$, a corresponding LEE can be generated on top of $|\psi_1\rangle$ using circuit $V' = UVU^{\dagger}$. Since $U$ is a locality preserving map, $V'$ has the same structure as $V$ (finite depth or sequential). Therefore, we conclude that the classification of LEEs is the same for $|\psi_0\rangle$ and $|\psi_1\rangle$. However, the parity defect considered here changes the locality of the system. Therefore, the QCA mapping between the states before the LEE is generated ($U$) and the one after the LEE is generated ($\tilde{U}$) are different. Therefore, $V' = \tilde{U}VU^{\dagger}$ does not necessarily have the same structure as $V$. Indeed, in the trivial phase given by $H_0 = -\sum_v X_v$, no action is needed to move the Hamiltonian from torus to Klein bottle. For the nontrivial phase, we need a 1d sequential circuit as shown below. 

As shown in Fig.~\ref{fig:DSt}, to get to a Klein bottle starting from a 2d torus, we cut open a boundary along a non-trivial cycle, flip one side of the boundary upside down, and reconnect back. Consider first the trivial SPT Hamiltonian $H_0 = -\sum_v X_v$. The ground state is a product state. The wavefunction does not change at all when the system is moved from a torus to Klein bottle. Of course, in the process, some degrees of freedom need to be moved around. We will assume that the flipping step that moves the degrees of freedom takes unit time so the whole process can be done in finite time. If we start from a slightly deformed Hamiltonian but still in the same trivial SPT phase, the process still needs only finite time because we can deform the Hamiltonian around the cut to the decoupled form ($-\sum_v X_v$) in finite time, do the flipping and then deform back in finite time. That is, moving from the torus to Klein bottle takes finite time for the trivial SPT state.

If we have a nontrivial SPT state, we can again try to open up a boundary, but the boundary is either gapless or breaks the symmetry due to the nontrivial SPT order in the bulk. Consider the case where the boundary spontaneously breaks the $Z_2$ symmetry. It takes linear time and a sequential circuit to generate such a boundary due to its long-range correlation. Therefore, if we now put together the three steps of opening up a boundary, flipping one side of the boundary, and reconnecting back, the total time needed scales linearly with the length of the boundary. Therefore, the parity defect in the nontrivial $Z_2$ SPT phase is a nontrivial LEE that can only be generated with a sequential circuit.

A similar cutting and gluing procedure can also be used to define the Hamiltonian of the SPT phase on the Klein bottle. The non-trivial SPT Hamiltonian considered in Sec.~\ref{sec:SPT} is explicitly parity symmetric, so defining the Hamiltonian on the Klein bottle can be interpreted as inserting a parity twist defect which is implemented as taking the Hamiltonian terms across the defect line and conjugating half of each by the parity symmetry. When the parity symmetry is emergent, it is less obvious that the Hamiltonian can still be consistently defined on the Klein bottle. To see that this is indeed the case, consider the cutting and gluing procedure above. In the cutting step, the Hamiltonian terms across the cut are tuned to zero, exposing the SPT edge states on the two side. Then a flip is performed on one side before coupling terms are added back to remove the edge states on the two sides. It is understood that, since after the flipping, the SPT order on the two sides remain the same, the edge states on the two sides can be coupled and removed, giving rise to the gapped Hamiltonian on the Klein bottle. This holds even when the parity symmetry of the Hamiltonian is explicitly broken.

Note that if we start with a nontrivial SPT state without emergent parity symmetry (with symmetry groups other than $Z_2$) and apply the procedure above, we can open a symmetry breaking boundary, do the flip but we cannot reconnect back into a non-symmetry breaking ground state on the whole Klein bottle. The parity defect remains symmetry-breaking, and is hence not a new type of defect beyond what is possible in a trivial SPT state.

\subsection{Realization in the Symmetry TFT formalism}

\begin{figure}[h]
    \centering
    \includegraphics[width= 0.8\textwidth]{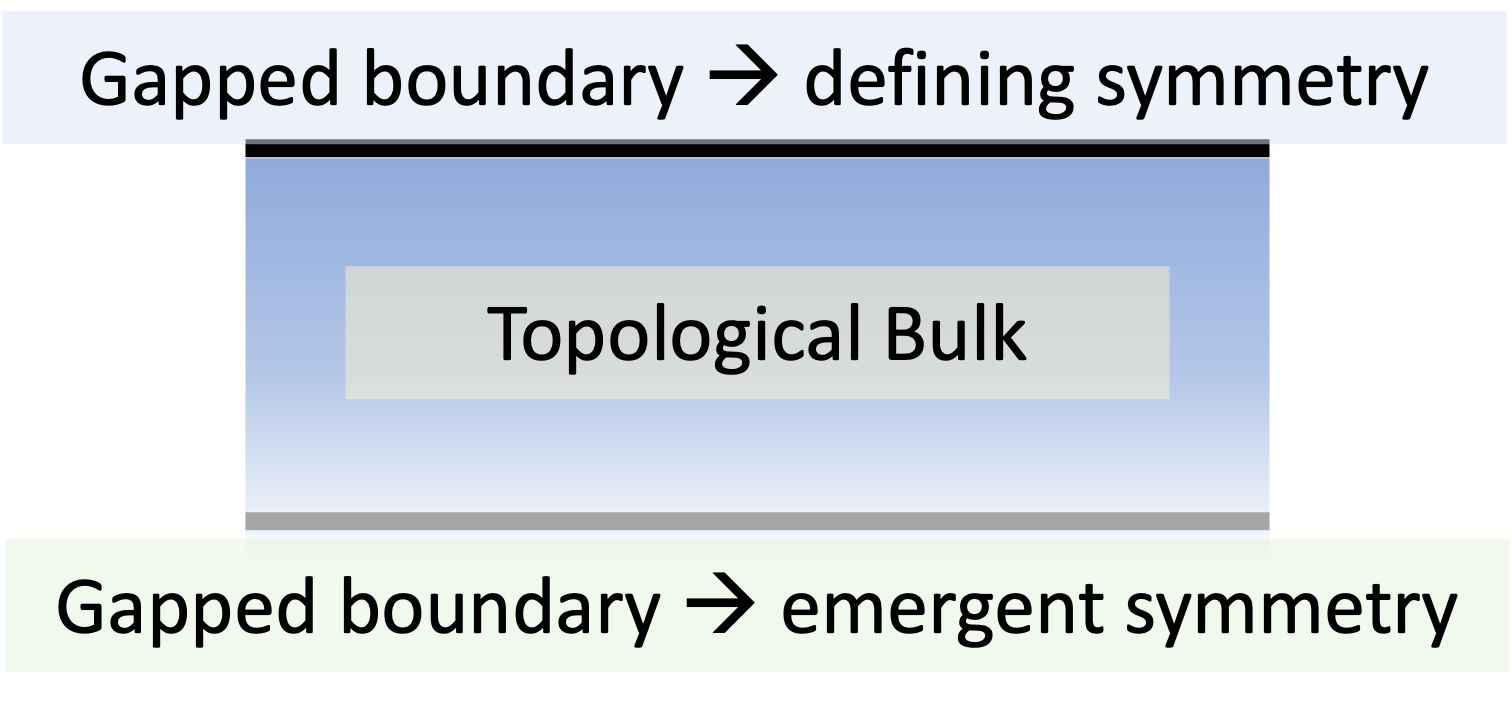}
    \caption{In the sandwich structure, the defining symmetry is determined by the top gapped boundary while the emergent symmetry is determined by the bottom gapped boundary.}
    \label{fig:SymTFTsym}
\end{figure}

In the symmetry TFT formalism, LEEs correspond to fractional excitations near the bottom boundary. Using the categorical notation in Ref.~\cite{Kitaev_2012}, they are denoted as $\mathcal{C}^*_{\mathcal{M}}$ in a $1+1$D sandwich, where $\mathcal{C}$ labels the category that defines the $2+1$D topological bulk and $\mathcal{M}$ is the module category that describes the gapped bottom boundary. Note that, the defining symmetry in the symmetry TFT formalism comes from the top boundary and is given by $\mathcal{C}^*_{\mathcal{M}'}$, where $\mathcal{M}'$ labels the module category of the top boundary. $\mathcal{C}^*_{\mathcal{M}}$ of the bottom boundary are emergent symmetries of the gapped phase. Since the dynamics of the system is controlled by the bottom boundary, emergent symmetries actually describe the dynamical properties of the system while $\mathcal{C}^*_{\mathcal{M}'}$ of the top boundary describe the kinetic properties of the system. In particular, when phase transitions happen, emergent symmetries in two phases compete with each other and give rise to all the complex critical phenomena (like universal scaling functions) while the top boundary only controls a few kinetic parameters like modular invariance of symmetry sectors. 

Realizing the same phase using different Symmetry TFT constructions can result in different defining symmetries for the phase and, correspondingly, different emergent symmetries. Consider, for example, the $Z_2$ topological ordered phase of $2+1$D Toric Code. It can be realized in the Symmetry TFT formalism with the $3+1$D Toric Code in the bulk and the smooth boundary condition at both the top and the bottom boundaries. Realized this way, a 1-form symmetry is enforced as the defining symmetry by the top boundary. The defect of the defining symmetry are point-like gauge charge excitations near the top boundary. Since the bottom boundary is the same as the top boundary, the excitations near the bottom boundary are also point-like gauge charge excitations, which gives rise to the 1-form emergent symmetry of the system. Since the emergent symmetry matches the defining symmetry, the 1-form symmetry is fully broken. On the other hand, we can choose to realize the $2+1$D Toric Code with a trivial $3+1$D bulk, a trivial top boundary and a $2+1$D layer of Toric Code at the bottom boundary. In this case, the defining symmetry is trivial. All LEE excitations of the $2+1$D Toric Code, including both gauge charge and gauge flux excitations, their descendant line excitations and the invertible defect line that exchanges gauge charge and gauge flux, lead to nontrivial emergent symmetries of the Toric Code. 

\section{Summary}

In this paper, we identify a nontrivial symmetry response of the $2+1$D $Z_2$ symmetry-protected topological phase when put on Klein bottle. In particular, inserting a symmetry defect line along the orientation reversing cycle of the Klein bottle induces an extra symmetry charge in the ground state. At the transition point between the trivial and nontrivial $Z_2$ SPT phases, the symmetry response results in an exact two-fold degeneracy in the ground state on the Klein bottle with the defect line inserted. We show that this result holds not only for the $Z_2$ SPT phase, but for all nontrivial SPT phases of order $2$ with abelian symmetries and type I and type II cocycles.

Inspired by this SPT model, especially the role played by parity symmetry, we explored in-depth the notion of emergent symmetry in gapped phases. We defined it using the tool of quantum circuit and discussed how it manifests in the Symmetry TFT framework. It will be interesting to explore how much of the discussion can be extended to gapless / critical systems. That is, to identify emergent symmetries in gapless or critical systems and to understand how that is related to the emergent symmetries in the neighboring gapped phases.


\begin{acknowledgments}
V.R. and X.C. are grateful for inspiring discussions with Nat Tantivasadakarn, Ruben Verresen, Robijn Vanhove, Wenjie Ji, and Xiao-Gang Wen.  V.R. is supported by the National Science Foundation Graduate Research Fellowship under Grant No. 2139433. 
X.C. is supported by the Simons collaboration on `Ultra-Quantum Matter'' (grant number 651438), the Simons Investigator Award (award ID 828078), the Institute for Quantum Information and Matter at Caltech (grant number PHY-2317110), and the Walter Burke Institute for Theoretical Physics at Caltech. 
B.Y. gratefully acknowledges support from Harvard CMSA and the Simons Foundation through Simons
Collaboration on Global Categorical Symmetries. 
\end{acknowledgments}
\newpage

\bibliography{references}

\appendix

\section{Calculating the SPT ground state symmetry charge using the gauged theory}
\label{app:gauge}
An independent calculation of the ground state symmetry charge calculated in Section.~\ref{sec:SPT} can be performed by gauging the global $Z_2$ symmetry. It is known that the result of gauging the non-trivial $Z_2$ SPT is the doubled semion (DS) topological order or twisted $Z_2$ gauge theory \cite{Levin_2005}. The DS model can be defined on the hexagonal lattice dual to the triangular lattice considered above and its Hamiltonian is comprised of vertex and plaquette terms, defined below. The global symmetry is promoted to a $Z_2$ gauge symmetry which is implemented by the vertex terms:\begin{equation}
    A_v = \vertex
\end{equation}

Under gauging, the SPT Hamiltonian terms $B_v$ map to plaquette terms:\begin{equation}
    B_p^{\text{DS}} = \DS
\end{equation}
with Pauli $X$ acting on the links around the plaquette $p$ and phase gates $S = \text{diag}(1,i)$ acting on the outgoing legs. These plaquette terms commute with each other in the gauge invariant subspace, i.e. states in which all the vertex terms are equal to 1. An excitation of the plaquette terms is a bosonic gauge charge denoted as $e$, and created in pairs by the string operator $\prod_{L^{\vee}} Z$ living on links of the dual lattice denoted as $L^{\vee}$. An excitation of the vertex terms on the other hand, can be created by two distinct string operators which are designed to commute with the plaquette terms. These excitations are gauge fluxes, labelled as semion $s$ (with topological spin $i$) and anti-semion $\bar{s}$ (with topological spin $-i$). The anti-semion is a dyon comprised of the semion and the gauge charge, i.e. $\bar{s} = es$.\\

Let us consider the action of gauging on the $Z_2$ SPT ground states on the Klein bottle. The symmetry uncharged ground states (no flux and flux along the orientable direction) are mapped into two degenerate ground states of DS. However, a $-1$ symmetry charge under gauging becomes a violation of vertex terms $A_v$, so the symmetry charged ground state must be mapped to excited states with a gauge charge. Hence on the Klein bottle we expect that among all 4 flux sectors, only two map to ground states. In order to show this, we prove that the product of all the plaquette terms on a Klein bottle is a Wilson line for $e$ around an orientable non-contractible loop $C$ on the dual lattice:
\begin{equation}\label{product}
    \prod_iB_{p_i}^{\text{DS}} = \prod_{j\in C}Z_j
\end{equation} 
The value of the line operator, which measures flux through the non-orientable direction, is fixed to 1 in a ground state and the GSD is reduced from 4 (on the torus) to 2 \cite{Chan_2016}. 

\begin{figure*}[t]
    \begin{center}
        \includegraphics[width=0.99\linewidth]{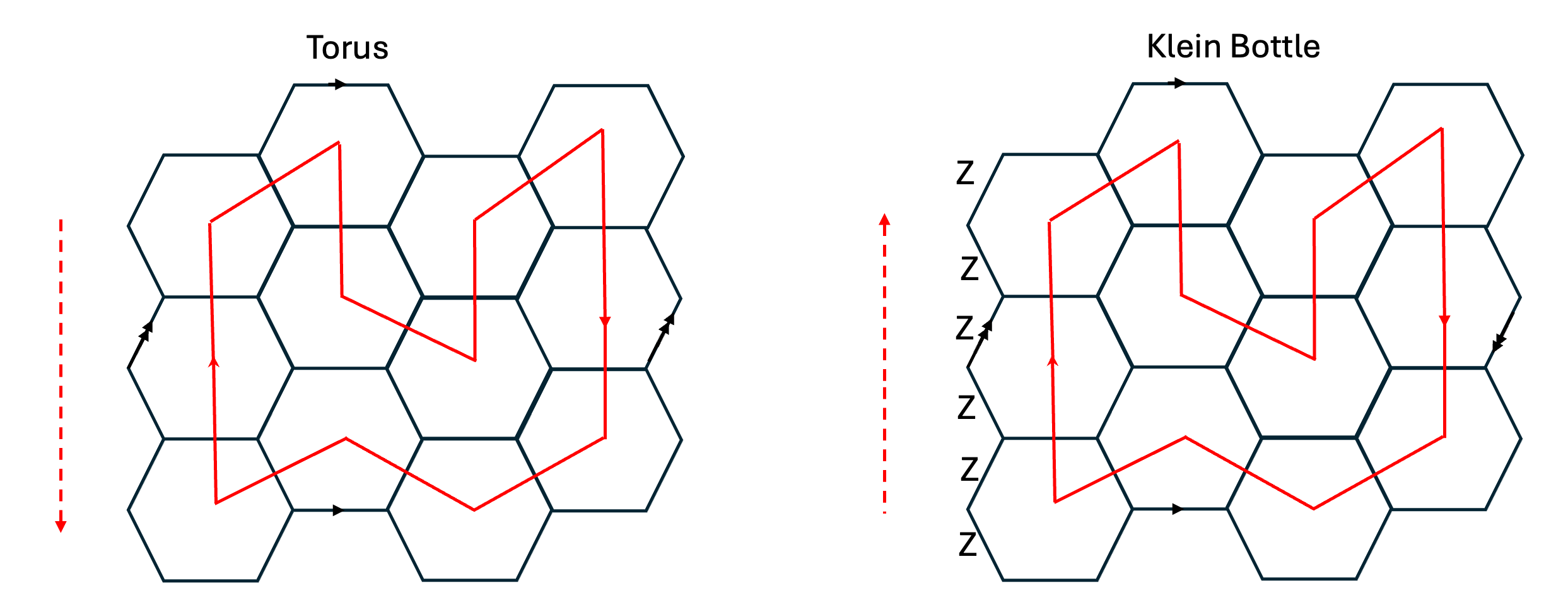}
    \end{center}
    \caption{Possible ordering of the non-commuting product $\prod_pB_p$ on the Torus (left) and Klein bottle (right). The red line passes through each plaquette exactly once and its orientation is the order in which the plaquettes are multiplied}
\label{fig:prod}
\end{figure*}

Let's calculate the product over all plaquette terms on a hexagonal lattice with torus and Klein bottle boundary conditions. The $B_p$ Hamiltonian terms of the model commute with each other in the gauge-invariant subspace, i.e. on states in which all the vertex terms $A_v=1$. All ground states must satisfy this condition, hence we are free to choose any ordering to evaluate the product $\prod_pB_p$. The ordering is chosen by drawing an oriented contractible loop on the dual lattice passing through every plaquette on the lattice once such as shown in Fig.~\ref{fig:prod}. The product is evaluated starting at any plaquette and following the loop till all the terms are multiplied. The result of the product on the link highlighted in green in the equations will depend on the configuration of the chosen loop surrounding it as follows:
\begin{equation}\label{config1}
    \begin{tikzpicture}[yshift = -15mm]
	\begin{pgfonlayer}{nodelayer}
		\node [style=none] (0) at (-2.5, 2.25) {};
		\node [style=none] (1) at (-3.25, 1.5) {};
		\node [style=none] (2) at (-2.5, 0.75) {};
		\node [style=none] (3) at (-1.5, 0.75) {};
		\node [style=none] (4) at (-0.75, 1.5) {};
		\node [style=none] (5) at (-1.5, 2.25) {};
		\node [style=none] (6) at (0.25, 1.5) {};
		\node [style=none] (7) at (-0.75, 3) {};
		\node [style=none] (8) at (0.25, 3) {};
		\node [style=none] (9) at (1, 2.25) {};
		\node [style=none] (10) at (-3.25, 3) {};
		\node [style=none] (11) at (-2.5, 3.75) {};
		\node [style=none] (12) at (-1.5, 3.75) {};
		\node [style=none] (13) at (-0.75, 0) {};
		\node [style=none] (14) at (0.25, 0) {};
		\node [style=none] (15) at (1, 0.75) {};
		\node [style=none] (16) at (-2, 3) {};
		\node [style=none] (17) at (-2, 1.5) {};
		\node [style=none] (18) at (-0.25, 2.25) {};
		\node [style=none] (19) at (-0.25, 0.75) {};
		\node [style=none] (20) at (-2, 1.5) {};
		\node [style=none] (22) at (-2, 1.5) {};
	\end{pgfonlayer}
	\begin{pgfonlayer}{edgelayer}
		\draw (0.center) to (5.center);
		\draw (4.center) to (3.center);
		\draw (3.center) to (2.center);
		\draw (2.center) to (1.center);
		\draw (1.center) to (0.center);
		\draw (10.center) to (0.center);
		\draw (10.center) to (11.center);
		\draw (11.center) to (12.center);
		\draw (12.center) to (7.center);
		\draw (7.center) to (5.center);
		\draw (7.center) to (8.center);
		\draw (8.center) to (9.center);
		\draw (9.center) to (6.center);
		\draw (6.center) to (4.center);
		\draw (3.center) to (13.center);
		\draw (13.center) to (14.center);
		\draw (14.center) to (15.center);
		\draw (15.center) to (6.center);
		\draw [style=green] (16.center) to (22.center);
		\draw [style=green] (18.center) to (19.center);
		\draw [style=gree] (5.center) to (4.center);
	\end{pgfonlayer}
\end{tikzpicture}
 = SXXS = Z
\end{equation}
\begin{equation}\label{config2}
  \begin{tikzpicture}[yshift = -15mm]
	\begin{pgfonlayer}{nodelayer}
		\node [style=none] (0) at (-2.5, 2.25) {};
		\node [style=none] (1) at (-3.25, 1.5) {};
		\node [style=none] (2) at (-2.5, 0.75) {};
		\node [style=none] (3) at (-1.5, 0.75) {};
		\node [style=none] (4) at (-0.75, 1.5) {};
		\node [style=none] (5) at (-1.5, 2.25) {};
		\node [style=none] (6) at (0.25, 1.5) {};
		\node [style=none] (7) at (-0.75, 3) {};
		\node [style=none] (8) at (0.25, 3) {};
		\node [style=none] (9) at (1, 2.25) {};
		\node [style=none] (10) at (-3.25, 3) {};
		\node [style=none] (11) at (-2.5, 3.75) {};
		\node [style=none] (12) at (-1.5, 3.75) {};
		\node [style=none] (13) at (-0.75, 0) {};
		\node [style=none] (14) at (0.25, 0) {};
		\node [style=none] (15) at (1, 0.75) {};
		\node [style=none] (16) at (-2, 3) {};
		\node [style=none] (17) at (-2, 1.5) {};
		\node [style=none] (18) at (-0.25, 2.25) {};
		\node [style=none] (19) at (-0.25, 0.75) {};
		\node [style=none] (20) at (-2, 1.5) {};
		\node [style=none] (22) at (-2, 1.5) {};
	\end{pgfonlayer}
	\begin{pgfonlayer}{edgelayer}
		\draw (0.center) to (5.center);
		\draw (4.center) to (3.center);
		\draw (3.center) to (2.center);
		\draw (2.center) to (1.center);
		\draw (1.center) to (0.center);
		\draw (10.center) to (0.center);
		\draw (10.center) to (11.center);
		\draw (11.center) to (12.center);
		\draw (12.center) to (7.center);
		\draw (7.center) to (5.center);
		\draw (7.center) to (8.center);
		\draw (8.center) to (9.center);
		\draw (9.center) to (6.center);
		\draw (6.center) to (4.center);
		\draw (3.center) to (13.center);
		\draw (13.center) to (14.center);
		\draw (14.center) to (15.center);
		\draw (15.center) to (6.center);
		\draw [style=green] (16.center) to (22.center);
		\draw [style=gree] (5.center) to (4.center);
		\draw [style=green] (19.center) to (18.center);
	\end{pgfonlayer}
\end{tikzpicture} = SXSX = i
\end{equation}
\begin{equation}\label{config3}
  \begin{tikzpicture}[yshift = -15mm]
	\begin{pgfonlayer}{nodelayer}
		\node [style=none] (0) at (-2.5, 2.25) {};
		\node [style=none] (1) at (-3.25, 1.5) {};
		\node [style=none] (2) at (-2.5, 0.75) {};
		\node [style=none] (3) at (-1.5, 0.75) {};
		\node [style=none] (4) at (-0.75, 1.5) {};
		\node [style=none] (5) at (-1.5, 2.25) {};
		\node [style=none] (6) at (0.25, 1.5) {};
		\node [style=none] (7) at (-0.75, 3) {};
		\node [style=none] (8) at (0.25, 3) {};
		\node [style=none] (9) at (1, 2.25) {};
		\node [style=none] (10) at (-3.25, 3) {};
		\node [style=none] (11) at (-2.5, 3.75) {};
		\node [style=none] (12) at (-1.5, 3.75) {};
		\node [style=none] (13) at (-0.75, 0) {};
		\node [style=none] (14) at (0.25, 0) {};
		\node [style=none] (15) at (1, 0.75) {};
		\node [style=none] (16) at (-2, 3) {};
		\node [style=none] (17) at (-2, 1.5) {};
		\node [style=none] (18) at (-0.25, 2.25) {};
		\node [style=none] (19) at (-0.25, 0.75) {};
		\node [style=none] (20) at (-2, 1.5) {};
		\node [style=none] (22) at (-2, 1.5) {};
	\end{pgfonlayer}
	\begin{pgfonlayer}{edgelayer}
		\draw (0.center) to (5.center);
		\draw (4.center) to (3.center);
		\draw (3.center) to (2.center);
		\draw (2.center) to (1.center);
		\draw (1.center) to (0.center);
		\draw (10.center) to (0.center);
		\draw (10.center) to (11.center);
		\draw (11.center) to (12.center);
		\draw (12.center) to (7.center);
		\draw (7.center) to (5.center);
		\draw (7.center) to (8.center);
		\draw (8.center) to (9.center);
		\draw (9.center) to (6.center);
		\draw (6.center) to (4.center);
		\draw (3.center) to (13.center);
		\draw (13.center) to (14.center);
		\draw (14.center) to (15.center);
		\draw (15.center) to (6.center);
		\draw [style=green] (16.center) to (22.center);
		\draw [style=gree] (5.center) to (4.center);
		\draw [style=green] (22.center) to (18.center);
		\draw [style=green] (18.center) to (19.center);
	\end{pgfonlayer}
\end{tikzpicture}
 = XSSX = -Z
\end{equation}
According to Eq.~(\ref{config3}) there will be $Z$s left behind on the links intersected by the loop. Since the loop is contractible, this product of $Z$s can be written as a product of $A_v$ terms and evaluates to $1$ on the ground states. On the torus, all the other links will be in configuration (\ref{config2}) and hence will only contribute a phase factor to the product. However on a Klein bottle there will be a loop of links corresponding to (\ref{config1}) going around the orientation preserving cycle. Hence the total product, upto phase factors, will be trivial on the torus and a string of $Z$s on the Klein bottle and we have shown Eq.~\ref{product}.

\end{document}